\documentclass[a4paper,11pt]{article}

\usepackage{amssymb,amsmath,amsthm}
\usepackage{texdraw}
\usepackage{epic,eepic}
\usepackage{tabularx}
\usepackage{booktabs}
\usepackage{authblk}
\usepackage{breqn}
\usepackage{color}

\usepackage{multirow}
\usepackage[thinlines]{easytable}
\usepackage{longtable}
\usepackage{makecell}
\usepackage{boldline}
\usepackage{subcaption}
\usepackage{scrextend}
\deffootnote{1.6em}{1.6em}{\thefootnotemark.\enskip}

\newtheorem{thm}{Theorem}[section]
\newtheorem{proposition}[thm]{Proposition}
\newtheorem{corollary}[thm]{Corollary}

\newtheorem{remark}[thm]{Remark}

\newcommand{\specialcell}[2][l]{\begin{tabular}[#1]{@{}l@{}}#2\end{tabular}}

\makeatletter
\def\hlinewd#1{\noalign{\ifnum0=`}\fi\hrule \@height #1 \futurelet\reserved@a\@xhline} 
\newcommand{\thickhline}{\noalign{\ifnum 0=`}\fi\hrule height 1pt \futurelet\reserved@a\@xhline}
\newcolumntype{"}{@{\hskip\tabcolsep\vrule width 1pt\hskip\tabcolsep}}
\makeatother

\newcolumntype{[}{@{\vrule width 1pt\hspace{7.5pt}}}
\newcolumntype{]}{@{\hspace{10.5pt}\vrule width 1pt}}
\newcolumntype{!}{@{\hskip\tabcolsep\vrule width 1pt\hskip\tabcolsep}}

\begin{document}

\title{\vspace{-0.5cm}
\textbf{Group classification of charged particle motion in stationary electromagnetic fields}}
\author{N. Kallinikos}
\affil{Department of Physics, Aristotle University of Thessaloniki, GR-54124 Thessaloniki, Greece \tt{kallinikos@auth.gr}
}
\date{\today}

\maketitle



\begin{abstract}
In this paper we classify in terms of Lie point symmetries the three-dimensional nonrelativistic motion of charged particles in arbitrary time-independent electromagnetic fields. The classification is made on the ground of equivalence transformations, and, when the system is nonlinear and particularly for inhomogeneous and curved magnetic fields, it is also complete. Using the homogeneous Maxwell's equations as auxiliary conditions for consistency, in which case the system amounts to a Lagrangian of three degrees of freedom with velocity-dependent potentials, the equivalence group stays the same. Therefore, instead of the actual fields, the potentials are equally employed and their gauge invariance results in an infinite-dimensional equivalence algebra, which nevertheless projects to finite-dimensional symmetry algebras. Subsequently, optimal systems of equivalence subalgebras are obtained that lead to one-, two- and three-parameter extended symmetry groups, besides the obvious time translations. Finally, based on symmetries of Noether type, aspects of complete integrability are discussed, as well. 
\end{abstract}


\section{Introduction}

Symmetries of differential equations \cite{bluman1,bluman2,hydon,olver,stephani} are one of the most effective methods to study nonlinear systems. Their significance makes the equivalence problem and the related group classification problem, in particular, important accordingly. In this case, a class of differential equations, i.e. a system that includes arbitrary functions or parameters, is under investigation, instead of a particular system, i.e. a member of the class. The ultimate goal then is to classify the members of the class, meaning the arbitrary elements, in terms of the symmetry groups admitted. To this end, equivalence transformations are employed, defined as maps between solutions for different members of a class, whereas symmetry ones are maps between solutions of the same member. The theory of equivalence groups was first presented by Ovsiannikov \cite{ovsiannikov} given in a context very similar to Lie's framework for symmetry ones. Then later on Ibragimov and coworkers introduced the infinitesimal version of the equivalence condition with numerous applications; see for example \cite{ibragimov1}. For further reading, see \cite{lisle}, but also \cite{schwarz} for related aspects between symmetry and equivalence theory.

In this work, we give a full Lie point symmetry classification for the Newtonian motion of charged particles in inhomogeneous, curved and time-independent electromagnetic fields. The equations of motion are given by the Lorentz force law and the resulting three second-order, autonomous differential equations in vector form,
\begin{equation}
\label{lor}
\ddot{\boldsymbol x}=\dot{\boldsymbol x}\times\boldsymbol{B}\left(\boldsymbol{x}\right)+\boldsymbol{E}\left(\boldsymbol{x}\right),
\end{equation}
where $\boldsymbol x=\boldsymbol x(t)$ is the position of the particle as a function of the time $t$ with dot denoting derivation with respect to $t$, while $\boldsymbol B$ and $\boldsymbol E$ are smooth vector functions, expressing the magnetic and the electric field, respectively. The latter are arbitrary, apart perhaps from the restrictions given by the homogeneous Maxwell's equations
\begin{align}
\label{maxBE}
\nabla\cdot\boldsymbol{B}=0,\quad\nabla\times\boldsymbol{E}=0\,.
\end{align}
In light of (\ref{maxBE}), equations (\ref{lor}) is an Euler-Lagrange system coming from a Lagrangian function with velocity-dependent generalized potentials,
\begin{equation}
\label{L}
L\left(\boldsymbol{x},\dot{\boldsymbol{x}}\right)=\frac{1}{2}\,\dot{\boldsymbol{x}}^2+\dot{\boldsymbol{x}}\cdot\boldsymbol{A}\left(\boldsymbol{x}\right)-\Phi(\boldsymbol{x}),
\end{equation}
where $\boldsymbol{A}$ is a vector function and $\Phi$ a scalar one known as the magnetic and the electric potential in the sense of $\boldsymbol{B}=\nabla\times\boldsymbol{A}$ and $\boldsymbol{E}=-\nabla\Phi$, respectively. 

Scalar second-order ordinary differential equations (ODEs) of the general (solved) form $\ddot{x}=f(t,x,\dot{x})$ have been fully classified in terms of symmetries\footnote{In this paper, by ``symmetry'' we will always mean ``Lie point symmetry'', even when not explicitly stated.} in \cite{mahomedleach1}, following earlier works of Lie himself for the case $\ddot{x}=f(t,x)$. Naturally, when the investigation moved to higher dimensions, it also adopted this order and started in the absence of first-order derivatives.

A first account of a symmetry classification of such equations in two dimensions was presented in \cite{sen} for autonomous systems $\ddot{\boldsymbol{x}}=\boldsymbol{f}(\boldsymbol{x})$ considering the conservative case. It was given more elaborately though in \cite{dam-sop2}, where symmetry algebras of all possible dimensions were obtained besides the corresponding potentials. For the three-dimensional case, which recovers equations (\ref{lor}) when $\boldsymbol{B}=\boldsymbol{0}$, a symmetry classification, although not full, has been demonstrated in \cite{dam-sop3a} in terms of characteristic examples for symmetry algebras of various dimensions, and then given completely in \cite{dam-sop3b} in terms of Noether symmetries. In \cite{tsa-pa1,tsa-pa2}, using an alternative method based on the geometry of the space considered, the authors reexamined these two problems, respectively, for Lie algebras generating at least one more symmetry apart from time translations. Regardless including non-conservative systems as well, their study showed that a few cases were missing from the previous results. For the non-autonomous case $\ddot{\boldsymbol{x}}=\boldsymbol{f}(t,\boldsymbol{x})$ either in two or three dimensions, a full symmetry classification is still missing, although the equivalence groups for these two classes have been recently found in \cite{meleshko2,meleshko3}, respectively. There, the general forms of systems admitting at least one symmetry have also been given, as well as for their linear counterpart admitting another symmetry besides the ones coming from the linearity of the system.

On the other hand, when first-order derivatives are present, systems of second-order ODEs have been classified in terms of symmetries only in the linear case with constant coefficients too by reducing them first to the form $\ddot{\boldsymbol{x}}=A\boldsymbol{x}$, $A$ being a constant matrix, with no derivatives again. See \cite{wafo} for the two- and three-dimensional cases (with some classes missing though) and \cite{campo} for the four-dimensional one (with partial corrections to \cite{wafo}). Generalizations to arbitrary dimensions have so far considered the structure of the maximal symmetry algebra as in \cite{boyko} and linearization criteria (see references therein).

In a previous work \cite{me}, we have presented a symmetry analysis for (\ref{lor}), excluding straight magnetic fields, which had been earlier studied \cite{ha-go1,ha-go2}, and homogeneous ones, since they belong to the linear cases we comment above. Starting there directly from the infinitesimal symmetry criterion, the general form of at least one more Lie point symmetry was found for the system, besides the obvious time translations. At the same time, the general form of the corresponding electromagnetic potentials was also obtained by solving the restrictions from the determining equations for $\boldsymbol{A}$ and $\Phi$ in full generality. The solutions were given in five general characteristic cases (which are actually ten considering subcases for the electric potential), the rest of which can be constructed by cyclic permutation of the indices.

Here, exploiting now the equivalence group, this study is extended to cases with a third or more symmetries and reach a complete classification for this class (excluding the aforementioned two-dimensional and linear cases) that allows exploring integrable cases, as well, at least in terms of Noether symmetries. In the process, the outcome of the symmetry analysis is also made more applicable, the reason being twofold. Although all possible cases with at least a second symmetry can be obtained from the previous results, they would still comprise a long list of complicated expressions too. Thus, first, determining whether the system for a given field admits or not any extra symmetries is still not an easy task. Then, secondly, moving on to systems with a third or more symmetries, the list of all the cases involved would be vast. Both these obstacles are overcome here with the aid of equivalence transformations, which are used to identify systems that are mapped to each other and therefore grouped together. In this way, all the symmetry groups admitted are separated into fewer disjoint classes described by the potentials in a representative form as simple as possible too. The investigation of the cases with at least a third symmetry is necessary in order to detect completely integrable systems of this class described at the end of this paper.

More specifically, in the present work, we find first the equivalence group for the class of differential equations (\ref{lor}), showing that it remains the same, when (\ref{maxBE}) are also taken into account. Thus, the potentials are equally used for the equivalence problem for more than one reason\,: unavoidably for expressing the solutions of the auxiliary equations (\ref{maxBE}) that will be needed in the classification procedure, preferably for reducing the arbitrary functions from six to four, and necessarily for considering Noether symmetries, as well. However, due to the gauge invariance of the magnetic potential, this leads to an infinite-dimensional algebra of the equivalence group. Nonetheless, the maximal symmetry algebra included is finite-dimensional and is, in fact, the one previously found from the symmetry criterion for nonlinear systems and curved magnetic fields. Therefore, up to equivalence transformations we completely classify this class in terms of Lie point symmetries. The classification follows from the decomposition of the equivalence algebra into optimal systems of disjoint subalgebras. This is made under the action of the adjoint group that consists of inner automorphisms of the equivalence algebra, i.e. equivalence transformations that map isomorphically equivalence subalgebras to one another. In this way, we separate it in one-, two- and three-dimensional subalgebras that project to symmetry algebras of the same dimension leaving aside the principal Lie algebra of symmetry generators that are admitted by all members of the class. Accordingly, we deduce when the system admits one-, two- and three-parameter extended symmetry groups apart from the obvious time translations. Finally, a Noether point symmetry classification follows easily from thereof and conclusions on complete integrability are drawn.

This paper is organized as follows. In section \ref{equivalencefields} we find the equivalence group for class (\ref{lor}) with and without the auxiliary conditions (\ref{maxBE}), proving that it stays the same. Next in section \ref{equivalencepotentials}, the equivalence group for the same class is found in terms of the potentials, and the resulting classifying equations that express the restrictions of the arbitrary functions in terms of the symmetries are displayed in section \ref{classifying}. Section \ref{optimal} is the core of this paper, where the optimal systems of equivalence subalgebras up to three dimensions are derived. Then section \ref{symmetryclasses} shows the results of the Lie point symmetry classification, and finally section \ref{noetherclasses} the same for Noether point symmetries in terms of which cases of complete integrability are classified.



\section{Equivalence transformations in terms of fields}\label{equivalencefields}

Using a Cartesian frame of reference, we start by rewriting the class of differential equations (\ref{lor}) enlarged by the auxiliary conditions expressing the time-independence of the fields,
\begin{subequations}
\label{lorentzsystemeq}
\begin{align}
\label{auxsystemeq}
B^i_t=E^i_t&=0\\
\label{primsystemeq}
\ddot{x}^i-\epsilon_{ijk}\dot{x}^jB^k-E^i&=0
\end{align}
\end{subequations}
Here $\epsilon_{ijk}$ is the Levi-Civita symbol and Einstein's summation convention has been adopted, assuming all indices from now on and throughout the rest of this paper take values from 1 to 3, unless stated otherwise. We consider augmented point transformations of the form
\begin{align}
\begin{split}
\label{augtrans}
\widetilde{t}&=\widetilde{t}\left(t,\boldsymbol{x}\,;\epsilon\right)=t+\epsilon\xi(t,\boldsymbol x)+O\!\left(\epsilon^2\right)\\
\widetilde{x}^{\,i}&=\boldsymbol{\widetilde{x}}\left(t,\boldsymbol{x}\,;\epsilon\right)=x^i+\epsilon\eta^i(t,\boldsymbol x)+O\!\left(\epsilon^2\right)\\
\widetilde{B}^{\,i}&=\widetilde{B}^i\left(t,\boldsymbol{x},\boldsymbol{B},\boldsymbol{E}\,;\epsilon\right)=B^i+\epsilon\lambda^i(t,\boldsymbol x,\boldsymbol{B},\boldsymbol{E})+O\!\left(\epsilon^2\right)\\
\widetilde{E}^{\,i}&=\widetilde{B}^i\left(t,\boldsymbol{x},\boldsymbol{B},\boldsymbol{E}\,;\epsilon\right)=E^i+\epsilon\sigma^i(t,\boldsymbol x,\boldsymbol{B},\boldsymbol{E})+O\!\left(\epsilon^2\right)
\end{split}
\end{align}
generated by the vector field
\begin{equation}
\label{generatoreqBE}
\textbf{V}=\xi(t,\boldsymbol x)\frac{\partial}{\partial t}+\eta^i(t,\boldsymbol x)\frac{\partial}{\partial x^i}+\lambda^i(t,\boldsymbol x,\boldsymbol B,\boldsymbol E)\frac{\partial}{\partial B^i}+\sigma^i(t,\boldsymbol x,\boldsymbol B,\boldsymbol E)\frac{\partial}{\partial E^i}.
\end{equation}
The equivalence condition then reads
\begin{align}
\label{eqc}
\begin{split}
\textbf{V}^{(2,1)}(B^i_t)=\textbf{V}^{(2,1)}(E^i_t)&=0\\
\textbf{V}^{(2,1)}(\ddot{x}^i-\epsilon_{ijk}\dot{x}^jB^k-E^i)&=0
\end{split}
\end{align}
whenever (\ref{lorentzsystemeq}) hold, where $\textbf{V}^{(2,1)}$ is the prolongation of $\textbf{V}$ up to second-order derivatives in terms of $x^i(t)$ and first-order ones in terms of $B^i(t,\boldsymbol{x})$ and $E^i(t,\boldsymbol{x}$). The notation and terminology adopted here stays close to \cite{lisle}.

After substitution of (\ref{lorentzsystemeq}), conditions (\ref{eqc}) for the auxiliary and the primary system take the form of polynomials in terms of the derivatives of both $x^i$ and $B^i$, $E^i$, as well. Therefore, as identities for all $t$, $x^i$, $B^i$, $E^i$, $\dot{x}^i$, $B^i_{x^j}$ and $E^i_{x^j}$ they break up into a larger set of partial differential equations (PDEs). The latter are the equations that determine the equivalence algebra, coming from the coefficients of the monomials in the first derivatives of $x^i$, $B^i$ and $E^i$, and reduce to\vspace{-0.35cm}
\begin{align}
\label{determiningeq2}
\begin{split}
\qquad\qquad\qquad\qquad\qquad\qquad\qquad\qquad\qquad\quad\xi_{tt}&=0\\
\eta^i_{x^jx^k}&=0\\
\xi_{x^i}=\eta^i_t&=0\\
\eta^i_{x^j}+\eta^j_{x^i}&=0\quad \text{for }i\neq j\\
\eta^i_{x^j}-\eta^l_{x^k}&=0\quad \text{for }i=j,\,l=k\\
-(\eta^j_{x^j}/3+\xi_t)B^i+\eta^i_{x^j}B^j-\lambda^i&=0\\
-2\xi_tE^i+\eta^i_{x^j}E^j-\sigma^i&=0
\end{split}
\end{align}
Their solution yields the general form of the equivalence generator $\textbf{V}$, from which we deduce that the group of equivalence transformations for the class (\ref{lorentzsystemeq}) is generated by a nine-dimensional Lie algebra spanned by the vector fields
\begin{align}
\label{eqal1}
\begin{split}
\textbf{V}_0&=\partial_t\\
\textbf{V}_1&=\partial_x\\
\textbf{V}_2&=\partial_y\\
\textbf{V}_3&=\partial_z\\
\textbf{V}_4&=x\partial_y-y\partial_x+B^1\partial_{B^2}-B^2\partial_{B^1}+E^1\partial_{E^2}-E^2\partial_{E^1}\\
\textbf{V}_5&=z\partial_x-x\partial_z+B^3\partial_{B^1}-B^1\partial_{B^3}+E^3\partial_{E^1}-E^1\partial_{E^3}\\
\textbf{V}_6&=y\partial_z-z\partial_y+B^2\partial_{B^3}-B^3\partial_{B^2}+E^2\partial_{E^3}-E^3\partial_{E^2}\\
\textbf{V}_7&=x\partial_x+y\partial_y+z\partial_z+E^1\partial_{E^1}+E^2\partial_{E^2}+E^3\partial_{E^3}\\
\textbf{V}_8&=t\partial_t-B^1\partial_{B^1}-B^2\partial_{B^2}-B^3\partial_{B^3}-2E^1\partial_{E^1}-2E^2\partial_{E^2}-2E^3\partial_{E^3}
\end{split}
\end{align}

\vspace{0.05cm}\begin{remark}\normalfont
The principal Lie algebra generating symmetries admitted by the primary system (\ref{primsystemeq}) for any $\boldsymbol B$ and $\boldsymbol E$ is just the well expected $\textbf{v}_0=\partial_t$.
\end{remark}
\begin{remark}\normalfont
The projections $\textbf{v}_i$ of $\textbf{V}_i$, $i=0,\ldots,8$  on the $(t,\boldsymbol x)$-space are none other than the symmetries found in \cite{me} for inhomogeneous and curved magnetic fields. Apart from $\textbf{v}_0$, they constitute the symmetry generator (see equation (21) therein)\vspace{-0.1cm}
\begin{equation}
\label{sym}
\textbf{v}=\sum\limits_{i=1}^8c_i\textbf{v}_i
\end{equation}

\vspace{-0.05cm}
\noindent $c_i$, $i=1,\ldots,8$ being arbitrary constants, which is a sum of three translations, three rotations and two scalings,\vspace{-0.3cm}
\begin{align}
\label{systemsymmetries}
\begin{split}
\qquad\quad~\textbf{v}_1&=\partial_x\\
\textbf{v}_2&=\partial_y\\
\textbf{v}_3&=\partial_z\\
\textbf{v}_4&=x\partial_y-y\partial_x\\
\textbf{v}_5&=z\partial_x-x\partial_z\\
\textbf{v}_6&=y\partial_z-z\partial_y\\
\textbf{v}_7&=x\partial_x+y\partial_y+z\partial_z\\
\textbf{v}_8&=t\partial_t
\end{split}
\end{align} 
\end{remark}


\subsection{The homogeneous Maxwell's equations as auxiliary conditions}

Now let us focus to electromagnetic fields respecting Maxwell's equations (\ref{maxBE}), as well. The latter can also be inserted as auxiliary equations, further enlarging the auxiliary system and thus restricting the original class (\ref{lorentzsystemeq}) to
\begin{subequations}
\label{lorentzmaxsystemeq}
\begin{align}
\label{auxsystemeq1}
\begin{split}
B^i_{x^i}&=0\\
\epsilon_{ijk}E^k_{x^j}&=0
\end{split}\\
\label{auxsystemeq2}
\begin{split}
B^i_t=E^i_t&=0
\end{split}\\
\label{primsystemeqmax}
\ddot{x}^i-\epsilon_{ijk}\dot{x}^jB^k-E^i&=0
\end{align}
\end{subequations}
Here, the equivalence condition for this class apart from (\ref{eqc}) now also includes
\begin{align}
\label{eqcmax}
\begin{split}
\textbf{V}^{(2,1)}(B^i_{x^i})&=0\\
\textbf{V}^{(2,1)}(\epsilon_{ijk}E^k_{x^j})&=0
\end{split}
\end{align}
whenever (\ref{auxsystemeq1})-(\ref{auxsystemeq2}) hold. Substituting (\ref{auxsystemeq1}) and (\ref{auxsystemeq2}), relations (\ref{eqcmax}) yield in turn additional equations, supplementary to the original ones (\ref{determiningeq2}) and interestingly enough without altering them,
\begin{align}
\begin{split}
\lambda^i_{x^i}&=0\\
\lambda^i_{B^j}-\eta^i_{x^j}&=0\quad\text{for }~i\neq j\\
\lambda^i_{B^j}-\eta^i_{x^j}-\lambda^l_{B^k}+\eta^l_{x^k}&=0\quad\text{for }~i=j,\,l=k\\
\lambda^i_{E^j}-\lambda^j_{E^i}&=0\quad\text{for }~i\neq j\\
\lambda^i_{E^j}&=0\quad \text{for }~i=j\\
\sigma^i_{x^j}-\sigma^j_{x^i}&=0\quad\text{for }~i\neq j\\
\sigma^i_{B^j}&=0\\
\sigma^i_{E^j}+\eta^j_{x^i}&=0\quad\text{for }~i\neq j\\
\sigma^i_{E^j}+\eta^j_{x^i}-\sigma^l_{E^k}-\eta^k_{x^l}&=0\quad\text{for }~i=j,\,l=k
\end{split}
\end{align}

\vspace{0.1cm}\noindent It is easily verified that the latter satisfy the solution of the former without any further restrictions whatsoever. In other words, we conclude that\,: 

\begin{proposition}\label{eqmax}\normalfont
The class of differential equations (\ref{lorentzsystemeq}) admits the same equivalence group with the subclass (\ref{lorentzmaxsystemeq}).
\end{proposition}

Recalling that equivalence transformations of a class are symmetry ones for the auxiliary system, note that in agreement with the above proposition $\textbf{V}_i$, $i=0,\ldots,6$ are indeed well-known symmetry generators of Maxwell's equations altogether in vacuum that belong to the Lie algebra of the Lorentz group (see for example \cite{ibragimovmax}). However, that system only admits the linear combination $\textbf{V}_7+\textbf{V}_8$, contrary to (\ref{maxBE}) which as we can see admits both $\textbf{V}_7$ and $\textbf{V}_8$ separately.

On the other hand, Proposition \ref{eqmax} is rather unusual in the following sense\,: additional constraints may either lead to new equivalence transformations, known as conditional, or they can further restrict the equivalence generator resulting in an equivalence subgroup. From this point of view, it is worthy of noting that the equivalence group stayed the same.

\section{Equivalence transformations in terms of potentials}\label{equivalencepotentials}

Since Maxwell's equations (\ref{maxBE}) did not change the scenery, it is preferable to describe the equivalence problem for charged particle motion in terms of the potentials of the electromagnetic field. Therefore, the class (\ref{lorentzmaxsystemeq}) is equally expressed as
\begin{subequations}
\label{lorentzpotsystemeq}
\begin{align}
\label{auxpotsystemeq}
\begin{split}
A^i_t=0,\quad\Phi_t&=0
\end{split}\\
\label{primpotsystemeq}
\ddot{x}^i-\dot{x}^j(A^j_{x^i}-A^i_{x^j})+\Phi_{x^i}&=0
\end{align}
\end{subequations}
for which we consider augmented equivalence transformations generated by the vector field
\begin{equation}
\label{equivX}
\textbf{V}=\xi(t,\boldsymbol x)\frac{\partial}{\partial t}+\eta^i(t,\boldsymbol x)\frac{\partial}{\partial x^i}+\mu^i(t,\boldsymbol x,\boldsymbol A,\Phi)\frac{\partial}{\partial A^i}+\nu(t,\boldsymbol x,\boldsymbol A,\Phi)\frac{\partial}{\partial\Phi}.
\end{equation}
As previously, $\textbf{V}$ is prolonged again up to second-order derivatives for $x^i$ as functions of $t$ and first-order ones for $A^i$ and $\Phi$ as functions of $t$ and $\boldsymbol{x}$, and the equivalence condition is applied
\begin{align}
\label{eqcpot}
\begin{split}
\textbf{V}^{(2,1)}(A^i_t)=0,\quad\textbf{V}^{(2,1)}(\Phi_t)&=0\\
\textbf{V}^{(2,1)}(\ddot{x}^i-\dot{x}^j(A^j_{x^i}-A^i_{x^j})+\Phi_{x^i})&=0
\end{split}
\end{align}
whenever equations (\ref{lorentzpotsystemeq}) hold. 
After substitutions, the equations coming from (\ref{eqcpot}) that determine the equivalence generators are reduced to the system
\begin{align}
\label{potDE}
\begin{split}
\begin{aligned}
\xi_{tt}&=0\\
\eta^i_{x^jx^k}&=0\\
\xi_{x^i}=\eta^i_t&=0\\
\eta^i_{x^j}+\,\eta^j_{x^i}&=0\quad \text{for }i\neq j\\
\eta^i_{x^j}-\eta^l_{x^k}&=0\quad \text{for }i=j,\,l=k\\
\mu^i_t=\mu^i_\Phi&=0
\end{aligned}\qquad\quad
\begin{aligned}
\mu^i_{x^j}-\mu^j_{x^i}&=0\quad \text{for }i\neq j\\
\mu^i_{A^j}+\,\eta^j_{x^i}&=0\quad \text{for }i\neq j\\
\mu^i_{A^j}-\eta^i_{x^j}+\xi_t&=0\quad \text{for }i=j\\
\nu_t&=0\\
\nu_{x^i}=\nu_{A^i}&=0\\
\nu_\Phi-2\left(\eta^i_{x^j}-\xi_t\right)&=0\quad \text{for }i=j
\end{aligned}
\end{split}
\end{align}
From their solution, we deduce that the equivalence algebra for the class of differential equations (\ref{lorentzpotsystemeq}) is spanned by the vector fields
\begin{align}
\label{eqal2}
\begin{split}
\textbf{V}_0&=\partial_t\\
\textbf{V}_1&=\partial_x\\
\textbf{V}_2&=\partial_y\\
\textbf{V}_3&=\partial_z\\
\textbf{V}_4&=x\partial_y-y\partial_x+A^1\partial_{A^2}-A^2\partial_{A^1}\\
\textbf{V}_5&=z\partial_x-x\partial_z+A^3\partial_{A^1}-A^1\partial_{A^3}\\
\textbf{V}_6&=y\partial_z-z\partial_y+A^2\partial_{A^3}-A^3\partial_{A^2}\\
\textbf{V}_7&=x\partial_x+y\partial_y+z\partial_z+A^1\partial_{A^1}+A^2\partial_{A^2}+A^3\partial_{A^3}+2\Phi\partial_{\Phi}\\
\textbf{V}_8&=t\partial_t-A^1\partial_{A^1}-A^2\partial_{A^2}-A^3\partial_{A^3}-2\Phi\partial_{\Phi}\\
\textbf{V}_9&=\partial_{\Phi}\\
\textbf{V}_f&=\nabla f\cdot\partial_{\boldsymbol{A}}
\end{split}
\end{align}
consisting now of the ten-dimensional subalgebra generated by $\textbf{V}_i$, $i=0,\ldots,9$ and the infinite-dimensional one $\textbf{V}_f$, where $f$ is an arbitrary function of $\boldsymbol{x}$. We see once again that the symmetry generators included are the $\textbf{v}_i$, $i=0,\ldots,8$ listed in the previous section, while the principal Lie algebra is just $\textbf{v}_0$. Accordingly, we arrive at the following theorem.

\begin{thm}\normalfont
The equivalence group for the class of systems (\ref{lorentzpotsystemeq}) consists of the transformations
\begin{align}
\begin{split}
\label{group}
\widetilde{t}&=\epsilon_8t+\epsilon_0\\
\widetilde{\boldsymbol{x}}&=\epsilon_7R_1(\epsilon_6)R_2(\epsilon_5)R_3(\epsilon_4)\boldsymbol{x}+\boldsymbol{\epsilon}\\
\widetilde{\boldsymbol{A}}&=\epsilon_7\epsilon_8^{-1}R_1(\epsilon_6)R_2(\epsilon_5)R_3(\epsilon_4)\boldsymbol{A}+\nabla g\\
\widetilde{\Phi}&=\epsilon_7^2\epsilon_8^{-2}\Phi+\epsilon_9
\end{split}
\end{align}
where $\epsilon_i$, $i=0,\ldots,9$, are arbitrary constants for $\epsilon_{7,8}\neq0$ with $\boldsymbol{\epsilon}=(\epsilon_1,\epsilon_2,\epsilon_3)$, $g$ is an arbitrary function of $\boldsymbol{x}$, and $R_i$ are the rotation matrices around $x^i$, respectively.
\end{thm}

\begin{remark}\normalfont
Besides the well-known gauge equivalence, one also recognizes the linear part of the transformations used in \cite{me} to solve the constraints for the potentials (see (35), (38), and (40) therein) as a subgroup of (\ref{group}).
\end{remark}

\begin{remark}\normalfont
By first inspection, we also note the following discrete equivalence transformations
\begin{align}
\begin{split}
1.&\quad\widetilde{t}=-\,t,~\widetilde{\boldsymbol{x}}=-\,\boldsymbol{x}\\
2.&\quad\widetilde{t}=-\,t,~\widetilde{\boldsymbol{A}}=-\boldsymbol{A}\\
3.&\quad\widetilde{t}=-\,t,~\widetilde{x}^i=-\,x^i,~\widetilde{x}^j=-\,x^j,~\widetilde{A}^k=-\,A^k\\
4.&\quad\widetilde{x}^i=x^j,~\widetilde{x}^j=x^i,~\widetilde{A}^i=A^j,~\widetilde{A}^j=A^i
\end{split}
\end{align}
for $i\neq j\neq k$, the first one being a discrete symmetry for all members of the class. 
\end{remark}

\begin{remark}\normalfont
Inclusion of generalized equivalence transformations in the sense of \cite{meleshko1}, i.e. that allow the transformations on the $(t,\boldsymbol{x})$-space to depend on the arbitrary functions $\boldsymbol{A}$ and $\Phi$ too, does not add anything new to the group (\ref{group}). (The tedious calculations involved are here omitted.)
\end{remark}

\section{Classifying equations}\label{classifying}

Having found the equivalence transformations, we proceed with the symmetry analysis of the class (\ref{lorentzpotsystemeq}) for particular forms of the electromagnetic field. Consider the equivalence subalgebra $\mathfrak{g}$, generated by the equivalence generators (\ref{eqal2}) leaving aside the principal generator $\textbf{V}_0$, and the corresponding equivalence subgroup $G$. Let $\bar{\mathfrak{g}}$ denote the nine-dimensional subalgebra of $\mathfrak{g}$ generated by $\textbf{V}_i$, $i=1,\ldots,9$. Any element of $\mathfrak{g}$ is written as the linear combination
\begin{align}
\label{eq}
\textbf{V}=\overline{\textbf{V}}+\textbf{V}_f=\sum\limits_{i=1}^9c_i\textbf{V}_i+\textbf{V}_f
\end{align}
As previously noted, the projection $\textbf{v}$ of $\textbf{V}$ to the $(t,\boldsymbol x)$-space is the linear combination (\ref{sym}) of the symmetries $\textbf{v}_i$, $i=0,\ldots,8$ (\ref{systemsymmetries}).

Following \cite{ibragimov1,lisle}, the vector field $\textbf{v}$ is a symmetry generator for the primary system (\ref{primpotsystemeq}) for specific values $A^i=A^i(\boldsymbol x)$ and $\Phi=\Phi(\boldsymbol x)$ if and only if the latter functions are invariant solutions for the auxiliary system (\ref{auxpotsystemeq}) with respect to $\textbf{V}$, that is, $\textbf{V}\left(A^i-A^i(\boldsymbol x)\right)=0$ and $\textbf{V}\left(\Phi-\Phi(\boldsymbol x)\right)=0$\footnote{In fact, instead of $\textbf{V}$ it suffices to take the projection of $\textbf{V}$ to the $(\boldsymbol x,\boldsymbol{A},\Phi)$-space. Since, however, this would slightly simplify just $\textbf{V}_8$ by leaving out $t\partial_t$, in order to spare notation we continue to consider the whole of $\textbf{V}$.}. These conditions are expressed as
\begin{align}
\begin{split}
\label{APheq}
(c_7x^j-\epsilon_{jkl}c_{7-l}x^k+c_j)A^i_{x^j}&=~\,(c_7-c_8)A^i-\epsilon_{ijk}c_{7-k}A^j+f_{x^i},\\
(c_7x^i-\epsilon_{ijk}c_{7-k}x^j+c_i)\Phi_{x^i}&=2(c_7-c_8)\Phi\,+c_9
\end{split}
\end{align}
and since the symmetries of the equivalence group are the only symmetries for inhomogeneous and curved magnetic fields it is of no surprise that they are none other than the ones found in \cite{me} (see equations (26)-(27) therein) from the symmetry condition.

Of course, the general solution to the above system for the electromagnetic potential was given there in terms of the symmetry generator. In other words, $\boldsymbol{A}$ and $\Phi$ were described as general functions of the constants $c_i$, $i=1,\ldots,9$ that determine the general form of the symmetry generator (\ref{sym}). But in order to know given an electromagnetic field if and which symmetries are admitted, this treatment must be somewhat reversed. In a manner of speaking, here (\ref{APheq}) are solved for the symmetries and the potentials simultaneously. Ultimately we reach a classification of these solutions in terms of symmetries, which severely reduces the amount of labor to a few cases that correspond to systems that cannot be mapped to one another through an equivalence transformation (\ref{group}).

One way to begin with would be to classify the solutions of (\ref{APheq}) already found in \cite{me} under (\ref{group}), i.e. separate them into classes of electromagnetic fields that cannot be mapped to one another through (\ref{group}). Their highly-complicated form leaves quite a space for human error though. An alternative would be to make the classification first, before solving them. However, applying (\ref{group}) directly to (\ref{APheq}) and separating it to disjoint subcases still involves a great amount of work. Instead the most effective way is to take the equivalence algebra and decompose it into disjoint subalgebras under the action of the equivalence group, as described in \cite{ibragimov1,ovsiannikov}. Starting with one-dimensional subalgebras, we continue to higher-dimensional ones that provide us with cases of potentials corresponding to more than one additional symmetries, which could be very difficult to identify from the solutions of \cite{me}.

\section{Classification of equivalence subalgebras}\label{optimal}

In order to put the classification scheme into action, the adjoint representation (see \cite{olver,ovsiannikov}) of the equivalence group $G$ is found first, expressing the inner automorphisms of the equivalence algebra $\mathfrak{g}$. From the infinitesimal action of the adjoint group $\text{Ad}\,G$, that is, the mapping $\text{ad}\,\textbf{V}\left(\textbf{W}\right)=\left[\textbf{V},\textbf{W}\right]$ for any $\textbf{V}$ and $\textbf{W}$ in $\mathfrak{g}$, the generators of $\text{Ad}\,G$ are determined from the Lie brackets of the generators of $G$. For the finite-dimensional subalgebra $\bar{\mathfrak{g}}$ these are given in the next table, while for the infinite-dimensional one the commutator of any $\textbf{V}_g=\nabla g\cdot\partial_{\boldsymbol{A}}$ with $\textbf{V}$ is $\left[\textbf{V},\textbf{V}_g\right]=\nabla\left((c_7x^i-\epsilon_{ijk}c_{7-k}x^j+c_i)\,g_{x^i}+\left(c_8-2c_7\right)g\right)\cdot\partial_{\boldsymbol{A}}$ and lies again in the infinite-dimensional algebra.

\vspace{0.3cm}

\renewcommand\arraystretch{1.5}
\begin{table}[h]
\centering
\begin{tabular}{[c!ccccccccc]}
\thickhline
$\left[~~,~\,\right]$&~~$\textbf{V}_1$&~~$\textbf{V}_2$&~~$\textbf{V}_3$&~~$\textbf{V}_4$&~~$\textbf{V}_5$&~~$\textbf{V}_6$&~~$\textbf{V}_7$&~~$\textbf{V}_8$&~\,$\textbf{V}_9$\\
\thickhline
$\textbf{V}_1$ &~~0&~~0&~~0&~~$\textbf{V}_2$&$-\textbf{V}_3$&~~0&~~$\textbf{V}_1$&~0&0\\
$\textbf{V}_2$ &~~0&~~0&~~0&$-\textbf{V}_1$&~~0&~~$\textbf{V}_3$&~~$\textbf{V}_2$&~0&0\\
$\textbf{V}_3$ &~~0&~~0&~~0&~~0&$~~\textbf{V}_1$&$-\textbf{V}_2$&~~$\textbf{V}_3$&~0&0\\
$\textbf{V}_4$ &$-\textbf{V}_2$&~~$\textbf{V}_1$&~~0&~~0&$~~\textbf{V}_6$&$-\textbf{V}_5$&~~0&~0&0\\
$\textbf{V}_5$ &~~$\textbf{V}_3$&~~0&$-\textbf{V}_1$&$-\textbf{V}_6$&~~0&~~$\textbf{V}_4$&~~0&~0&0\\
$\textbf{V}_6$ &~~0&$-\textbf{V}_3$&~~$\textbf{V}_2$&~~$\textbf{V}_5$&$-\textbf{V}_4$&~~0&~~0&~0&0\\
$\textbf{V}_7$ &$-\textbf{V}_1$&$-\textbf{V}_2$&$-\textbf{V}_3$&~~0&~~0&~~0&~~0&~0&$-2\textbf{V}_9$\\
$\textbf{V}_8$ &~~0&~~0&~~0&~~0&~~0&~~0&~~0&~0&$~~2\textbf{V}_9$\\
$\textbf{V}_9$ &~~0&~~0&~~0&~~0&~~0&~~0&~$2\textbf{V}_9$&$-2\textbf{V}_9$&0\\
\thickhline
\end{tabular}
\caption{Lie brackets between the equivalence generators in $\bar{\mathfrak{g}}$.}
\label{brackets}
\end{table}
\renewcommand\arraystretch{1.0}

Following \cite{ibragimov1}, the adjoint action is considered on the constants $c_i$, $i=1,\ldots,9$ instead of the actual vector fields $\textbf{V}_i$, $i=1,\ldots,9$ that span the finite-dimensional algebra $\bar{\mathfrak{g}}$. The same idea is applied to the infinite-dimensional algebra and instead of $\textbf{V}_f$ the adjoint representation on the function $f$ is given. In this notion the adjoint transformations are\,:
\begin{subequations}
\label{automorphisms}
\begin{enumerate}
\item\quad $\widetilde{c}_1=c_1+\epsilon_1c_7,\qquad\widetilde{c}_2=c_2+\epsilon_1c_4,\qquad\widetilde{c}_3=c_3-\epsilon_1c_5$ \hfill\refstepcounter{equation}(\theequation)\label{auto1}
\item\quad $\widetilde{c}_1=c_1-\epsilon_2c_4,\qquad\widetilde{c}_2=c_2+\epsilon_2c_7,\qquad\widetilde{c}_3=c_3+\epsilon_2c_6$ \hfill\refstepcounter{equation}(\theequation)\label{auto2}
\item\quad $\widetilde{c}_1=c_1+\epsilon_3c_5,\qquad\widetilde{c}_2=c_2-\epsilon_3c_6,\qquad\widetilde{c}_3=c_3+\epsilon_3c_7$ \hfill\refstepcounter{equation}(\theequation)\label{auto3}
\item\quad $\widetilde{c}_1=c_1\cos\epsilon_4+c_2\sin\epsilon_4,\qquad\widetilde{c}_5=c_5\cos\epsilon_4-c_6\sin\epsilon_4,\\ ~~~~\widetilde{c}_2=c_2\cos\epsilon_4-c_1\sin\epsilon_4,\qquad\widetilde{c}_6=c_6\cos\epsilon_4+c_5\sin\epsilon_4$ \hfill\refstepcounter{equation}(\theequation)\label{auto4}
\item\quad $\widetilde{c}_1=c_1\cos\epsilon_5-c_3\sin\epsilon_5,\qquad\widetilde{c}_4=c_4\cos\epsilon_5+c_6\sin\epsilon_5,\\ ~~~~\widetilde{c}_3=c_3\cos\epsilon_5+c_1\sin\epsilon_5,\qquad\widetilde{c}_6=c_6\cos\epsilon_5-c_4\sin\epsilon_5$ \hfill\refstepcounter{equation}(\theequation)\label{auto5}
\item\quad $\widetilde{c}_2=c_2\cos\epsilon_6+c_3\sin\epsilon_6,\qquad\widetilde{c}_4=c_4\cos\epsilon_6-c_5\sin\epsilon_6,\\ ~~~~\widetilde{c}_3=c_3\cos\epsilon_6-c_2\sin\epsilon_6,\qquad\widetilde{c}_5=c_5\cos\epsilon_6+c_4\sin\epsilon_6$ \hfill\refstepcounter{equation}(\theequation)\label{auto6}
\item\quad $\widetilde{c}_1=\epsilon_7c_1,\qquad\widetilde{c}_2=\epsilon_7c_2,\quad~\,\,\,\widetilde{c}_3=\epsilon_7c_3$ \hfill\refstepcounter{equation}(\theequation)\label{auto7}
\item\quad $\widetilde{c}_9=\epsilon_8c_9$ \hfill\refstepcounter{equation}(\theequation)\label{auto8}
\item\quad $\widetilde{c}_9=\left(c_7-c_8\right)\!\epsilon_9+c_9$ \hfill\refstepcounter{equation}(\theequation)\label{auto9}
\item\quad $\widetilde{f}=f-(c_7x^i-\epsilon_{ijk}c_{7-k}x^j+c_i)\,g_{x^i}+\left(2c_7-c_8\right)g+\text{const.}$ \hfill\refstepcounter{equation}(\theequation)\label{autof}
\end{enumerate}
\end{subequations}
where $\epsilon_i$, $i=1,\ldots,9$ are arbitrary constants with $\epsilon_{7,8}\neq0$ and $g$ is an arbitrary function of $\boldsymbol{x}$. Invariants of the adjoint group action (\ref{automorphisms}) are $c_7$, $c_8$ and $c=c_4^2+c_5^2+c_6^2$.

Now, using (\ref{automorphisms}) the equivalence algebra $\mathfrak{g}$ generated by $\textbf{V}_i$, $i=1,\ldots,9$ and $\textbf{V}_f$ can be decomposed into optimal systems of $n$-dimensional subalgebras. The forthcoming classification is performed for $n=1,2,3$ given in terms of their generators, which are expressed as\vspace{-0.1cm}
\begin{equation}
\label{span}
\textbf{Y}_i=\overline{\textbf{Y}}_i+\textbf{V}_{f_i}=\sum\limits_{j=1}^9c_{ij}\textbf{V}_j+\textbf{V}_{f_i}
\end{equation}
for $i=1,\ldots,n$. The method for obtaining optimal systems of subalgebras is outlined in \cite{ibragimov1,ovsiannikov}. In the presentation that follows, the tedious calculations for the analysis of the finite-dimensional part of $\mathfrak{g}$ are avoided, and emphasis is given on the inclusion of the infinite-dimensional part that comes next. For this purpose, to each generator $\textbf{Y}_l$, $l=1,\ldots,n$ a related operator is also assigned, namely
\begin{equation}
\label{releq}
U_l=\overline{\textbf{Y}}_l+c_{l8}-2c_{l7}=(c_{l7}x^i-\epsilon_{ijk}c_{l,7-k}x^j+c_{li})\partial_{x^i}+c_{l8}-2c_{l7}
\end{equation}

\subsection{One-dimensional equivalence subalgebras}\label{op1d}

In order to derive the optimal system of one-dimensional equivalence subalgebras, the general equivalence generator $\textbf{V}$ (\ref{eq}) is considered and, employing transformations (\ref{automorphisms}), the parameters $c_i$, $i=1,\ldots,9$ and the function $f$ are eliminated as much as possible. The resulting set of $c_{1i}=\widetilde{c}_i$, $i=1,\ldots,9$ and $f_1=\widetilde{f}$ yield a generator $\textbf{Y}_1$ that cannot be reduced furthermore.

Starting with the finite-dimensional equivalence algebra $\bar{\mathfrak{g}}$, i.e. the operator $\overline{\textbf{V}}$, the elimination of $c_i$, $i=1,\ldots,9$ that leads to $c_{1i}$, $i=1,\ldots,9$, i.e. the reduced operator $\overline{\textbf{Y}}_1$, breaks down to subcases depending on the invariants $c$, $c_7$ and $c_8$. In this way the generators of the one-dimensional subalgebras listed in Table \ref{optimal1d} are found that cannot be mapped to one another through (\ref{automorphisms}). From (\ref{APheq}), note that the symmetry $\textbf{v}_8$ coming from $\textbf{V}_8$ is admitted by the system for zero $\boldsymbol{A}$ and $\Phi$. Moreover $\textbf{V}_9$ does not project to any symmetries at all. As a consequence $\textbf{V}_8$ and $\textbf{V}_9$ do not appear in the optimal system just by themselves. 

In all these cases, the infinite-dimensional algebra $\textbf{V}_f$ if included as in $\overline{\textbf{Y}}_1+\textbf{V}_f$, then it can always be removed, using (\ref{autof}) and a suitable function $g$ such that $f_1=\text{const.}$. The latter is always possible, since it results in $U_1g=f$, which as we can see from (\ref{releq}) is a first-order linear partial differential equation whose characteristic equations for the independent variables are also linear and independent of the dependent one. Therefore $\textbf{Y}_1=\overline{\textbf{Y}}_1$. On the other hand, $\textbf{V}_f$ themselves, which are the only remaining vector fields outside $\bar{\mathfrak{g}}$, are not projected to any symmetries. Consequently the classification of one-dimensional subalgebras under the adjoint group based on the finite-dimensional algebra $\bar{\mathfrak{g}}$ is essentially the same for the infinite-dimensional one $\mathfrak{g}$.

In conclusion, the optimal system of one-dimensional equivalence subalgebras comprises of the operators collected in Table \ref{optimal1d}, where $k$'s and $\lambda$'s are constants.


\renewcommand\arraystretch{1.3}
\begin{table}[ht]
\centering
\begin{tabular}{[c!ll]}
\thickhline
& \multicolumn{2}{c]}{Operator}  \\
\thickhline
~1~ & $\textbf{V}_4+k_1\textbf{V}_7+k_2\textbf{V}_8$, & $k_2\neq k_1\neq0$ \\
\hline
2 & $\textbf{V}_4+k\left(\textbf{V}_7+\textbf{V}_8+\lambda\textbf{V}_9\right)$, & $k\neq0$ \\
\hline
3 & $\textbf{V}_4+k_1\textbf{V}_3+k_2\textbf{V}_8$, & $k_2\neq0$ \\
\hline
4 & $\textbf{V}_4+k\textbf{V}_3+\lambda\textbf{V}_9$ &\\
\hline
5 & $\textbf{V}_7+k\textbf{V}_8$, & $k\neq1$ \\
\hline
6 & $\textbf{V}_7+\textbf{V}_8+\lambda\textbf{V}_9$ & \\
\hline
7 & $\textbf{V}_3+k\textbf{V}_8$, & $k\neq0$ \\
\hline
8 & $\textbf{V}_3+\lambda\textbf{V}_9$ &\\
\thickhline
\end{tabular}
\caption{Optimal system of one-dimensional equivalence subalgebras.}
\label{optimal1d}
\end{table}
\renewcommand\arraystretch{1.0}

\subsection{Two-dimensional equivalence subalgebras}

For the construction of the optimal system of two-dimensional equivalence subalgebras, one of the two basis vectors is considered to be an operator $\textbf{Y}_1$ from the optimal system of one-dimensional subalgebras. Starting with the general generator $\textbf{V}$ (\ref{eq}), then the other one is determined from the relation, 
\begin{equation}
\label{2sub}
\left[\textbf{Y}_1,\textbf{V}\right]=\alpha\textbf{Y}_1+\beta\textbf{V}
\end{equation}
where $\alpha$ and $\beta$ are constants. Just like before, any two-dimensional equivalence subalgebras of $\mathfrak{g}$ lying completely outside $\bar{\mathfrak{g}}$ will not project to two-dimensional symmetry algebras. Therefore inclusion of the infinite-dimensional algebra $\textbf{V}_f$ is only made in $\textbf{V}$. For the operator $\overline{\textbf{V}}$ of the finite-dimensional subalgebra $\bar{\mathfrak{g}}$, equation (\ref{2sub}) results in an algebraic system, from which $\alpha$, $\beta$ and the coefficients $c_i$, $i=1,\ldots,9$ of $\overline{\textbf{V}}$ are determined, while the parameters $c_{1i}$, $i=1,\ldots,9$ of $\textbf{Y}_1$ may be further restricted. From $\textbf{V}_f$ it leads to a differential equation for $f$,
\begin{equation}
\label{con2d1}
(U_1-\beta)f=l
\end{equation}
where $l$ is an arbitrary constant. As the operator $\textbf{Y}_1$ runs Table \ref{optimal1d}, all possible cases of two-dimensional equivalence subalgebras are obtained at this stage.

In order to arrive at the optimal system of two-dimensional subalgebras $\left\{\textbf{Y}_1,\textbf{Y}_2\right\}$ spanned by $\textbf{Y}_1$ and $\textbf{Y}_2$, transformations (\ref{automorphisms}) are used again to simplify $\textbf{V}$ as much as possible into an operator $\textbf{Y}_2$ without changing $\textbf{Y}_1$. Taking (\ref{auto1})-(\ref{auto9}), first the constants $c_i$, $i=1,\ldots,9$ are reduced to $c_{2i}=\widetilde{c}_i$, $i=1,\ldots,9$, meaning $\overline{\textbf{V}}$ is reduced to $\overline{\textbf{Y}}_2$. Then, under the inner automorphism (\ref{autof}) $\textbf{V}_f$ is reduced too from $\overline{\textbf{Y}}_2+\textbf{V}_f$. For this to be the case a function $g$ is required that leaves $f_1$ from $\textbf{Y}_1$ unaltered and at the same time replaces $f$ with a simpler form $f_2=\widetilde{f}$ resulting in the second basis vector $\textbf{Y}_2$. Accordingly $g$ must satisfy a system of two first-order linear PDEs,
\begin{align}
\label{con2d2}
\begin{split}
U_1g&=n_1\\
U_2g&=n_2+f-f_2
\end{split}
\end{align}
where $n_1$ and $n_2$ are arbitrary constants, $f$ is the general solution to (\ref{con2d1}) and $f_2$ can be any solution to (\ref{con2d1}). Note that at this point the structure constants $\alpha$ and $\beta$ as well as the reduced parameters $c_{1i}$ and $c_{2i}$, $i=1,\ldots,9$ that define $U_1$ and $U_2$, respectively, have already been found. 
The compatibility condition of (\ref{con2d2}) yields $(U_1-\beta)(f-f_2)=\widetilde{l}$, where $\widetilde{l}$ is always a constant expressed as $\widetilde{l}=n_1\alpha+n_2\beta+\epsilon_{ij3}n_i(c_{j8}-2c_{j7})$, and whose vanishing thus depends on both $\textbf{Y}_1$ and $\overline{\textbf{Y}}_2$ for each case. Therefore, for the two-dimensional subalgebras $\left\{\textbf{Y}_1,\textbf{Y}_2\right\}$ in the optimal system the function $f_2$ of the operator $\textbf{Y}_2$ can be chosen either as\,: $i)$ the trivial solution of the homogeneous counterpart of (\ref{con2d1}), i.e. for $l=0$, when $\widetilde{l}\neq0$ by setting $\widetilde{l}=l$, or $ii)$ any particular solution of (\ref{con2d1}) when $\widetilde{l}=0$. In the former case $\textbf{V}_f$ is completely removed, meaning $\textbf{Y}_2=\overline{\textbf{Y}}_2$, while in the latter the contribution of the infinite-dimensional algebra to the second basis vector $\textbf{Y}_2$ is nonzero.

In conclusion, the optimal system of two-dimensional equivalence subalgebras consists of the algebras listed in Table \ref{optimal2d}, where $k$'s and $\lambda$'s are constants. The role played by the infinite-dimensional algebra on the second basis vector can be seen in cases 3, 5, 8, 10 and 13 via the operators $\textbf{V}_\phi=\rho^{-2}(x\partial_{A^2}-y\partial_{A^1})$, $\textbf{V}_{\ln z}=z^{-1}\partial_{A^3}$ and $\textbf{V}_z=\partial_{A^3}$, where $\rho=\sqrt{x^2+y^2}$ and $\phi=\arctan(y/x)$ are polar coordinates in the $xy$-plane.

\vspace{0.3cm}

\renewcommand\arraystretch{1.3}
\begin{table}[ht]
\centering
\begin{tabular}{[c!lll]}
\thickhline
& \multicolumn{3}{c]}{Operators} \\
\thickhline
~1~ & $\textbf{V}_3$, & $\textbf{V}_4+k_1\textbf{V}_7+k_2\textbf{V}_8$, & $2k_2,k_2\neq k_1\neq0$\\
\hline
2 & $\textbf{V}_3+\lambda\textbf{V}_9$, & $\textbf{V}_4+k(2\textbf{V}_7+\textbf{V}_8)$, & $k\neq0$\\
\hline
3 & $\textbf{V}_3+\lambda_1\textbf{V}_\phi$, & $\textbf{V}_4+k\left(\textbf{V}_7+\textbf{V}_8+\lambda_2\textbf{V}_9\right)$, & $k\neq0$\\
\hline
4 & $\textbf{V}_3+k_1\textbf{V}_8$, & $\textbf{V}_4+k_2\textbf{V}_8$, & $k_1\neq0$ or $k_2\neq0$\\
\hline
5 & $\textbf{V}_3+\lambda_1\textbf{V}_9+\lambda_3\textbf{V}_\phi$, & $\textbf{V}_4+\lambda_2\textbf{V}_9$ &\\
\hline
6 & $\textbf{V}_4+k_1\textbf{V}_8$, & $\textbf{V}_7+k_2\textbf{V}_8$, & $k_1\neq0$ or $k_2\neq1,2$\\
\hline
7 & $\textbf{V}_4+\lambda_1\textbf{V}_9$, & $\textbf{V}_7+\textbf{V}_8+\lambda_2\textbf{V}_9$ &\\
\hline
8 & $\textbf{V}_4$, & $\textbf{V}_7+2\textbf{V}_8+\lambda\textbf{V}_\phi$ &\\
\hline
9 & $\textbf{V}_3$, & $\textbf{V}_7+k\textbf{V}_8$, & $k\neq1/2,1$\\
\hline
10 & $\textbf{V}_3-\lambda_1\textbf{V}_{\ln z}$, & $\textbf{V}_7+\textbf{V}_8+\lambda_2\textbf{V}_9$ &\\
\hline
11 & $\textbf{V}_3+\lambda\textbf{V}_9$, & $2\textbf{V}_7+\textbf{V}_8$ &\\
\hline
12 & $\textbf{V}_2+k_1\textbf{V}_8$, & $\textbf{V}_3+k_2\textbf{V}_8$, & $k_2\neq0$\\
\hline
13 & $\textbf{V}_2+\lambda_1\textbf{V}_9+\lambda_3\textbf{V}_z$, & $\textbf{V}_3+\lambda_2\textbf{V}_9$ &\\
\thickhline
\end{tabular}
\caption{Optimal system of two-dimensional equivalence subalgebras.}
\label{optimal2d}
\end{table}
\renewcommand\arraystretch{1.0}

\subsection{Three-dimensional equivalence subalgebras}

Subsequently the optimal system of three-dimensional subalgebras $\left\{\textbf{Y}_1,\textbf{Y}_2,\textbf{Y}_3\right\}$ spanned by $\textbf{Y}_1$, $\textbf{Y}_2$ and $\textbf{Y}_3$ is constructed. Three-dimensional equivalence subalgebras could follow accordingly from two-dimensional ones, the only exception being (any algebra isomorphic to) the Lie algebra $\mathfrak{so}(3)$ that generates the rotation group (see \cite{ovsiannikov}). In other words, besides the latter case, we consider three-dimensional algebras that contain a two-dimensional subalgebra $\left\{\textbf{Y}_1,\textbf{Y}_2\right\}$ of the optimal system in Table \ref{optimal2d}; in accordance with the previous subsections, $\textbf{Y}_1=\overline{\textbf{Y}}_1$ is taken as the operator for which $f_1=\text{const.}$, while $\textbf{Y}_2=\overline{\textbf{Y}}_2+\textbf{V}_{f_2}$ is the one for which $f_2$ may be non-constant. Necessary and sufficient conditions for the third basis vector, taken as the general generator $\textbf{V}$ (\ref{eq}), are
\begin{align}
\begin{split}
\label{3sub}
\left[\textbf{Y}_1,\textbf{V}\right]&=\alpha_1\textbf{Y}_1+\beta_1\textbf{Y}_2+\gamma_1\textbf{V}\\
\left[\textbf{Y}_2,\textbf{V}\right]&=\alpha_2\textbf{Y}_1+\beta_2\textbf{Y}_2+\gamma_2\textbf{V}
\end{split}
\end{align}
where $\alpha_1$, $\alpha_2$, $\beta_1$, $\beta_2$, $\gamma_1$ and $\gamma_2$ are constants. Same as before, on one hand, conditions (\ref{3sub}) result in an algebraic system in terms of $\alpha_i,\beta_i,\gamma_i$, $i=1,2$ and the coefficients $c_i$, $i=1,\ldots,9$ of $\overline{\textbf{V}}$, which is solved accordingly, that may also further restrict the parameters $c_{1i}$ and $c_{2i}$, $i=1,\ldots,9$ of $\textbf{Y}_1$ and $\textbf{Y}_2$, respectively. And then using the adjoint transformations (\ref{auto1})-(\ref{auto9}) the generator $\overline{\textbf{V}}$ is reduced as much as possible to an operator $\overline{\textbf{Y}}_3=c_{3i}\textbf{V}_i$ for $c_{3i}=\widetilde{c}_i$, $i=1,\ldots,9$ without changing either $\textbf{Y}_1$ or $\overline{\textbf{Y}}_2$.

In consideration of the infinite-dimensional algebra $\textbf{V}_f$ in $\textbf{V}$, on the other, equations (\ref{3sub}) lead to a system of first-order linear PDEs for $f$,
\begin{align}
\label{con3d1}
\begin{split}
(U_1-\gamma_1)f&=l_1+\beta_1f_2\\
(U_2-\gamma_2)f&=l_2+(U_3+\beta_2)f_2
\end{split}
\end{align}
where $l_1$ and $l_2$ are arbitrary constants. The compatibility condition of (\ref{con3d1}) always reduces to $Uf_2=m$, where $U=\left(\alpha_1+\beta_2+U_3\right)U_1$, $m=l_1\alpha+l_2\beta+\epsilon_{ij3}l_i(c_{j8}-2c_{j7}-\gamma_j)$ and $\alpha$ and $\beta$ are the structure constants of the two-dimensional subalgebra $\left\{\textbf{Y}_1,\textbf{Y}_2\right\}$ included, which are already known. This equation may further restrict or completely eliminate the parameters $l_i$, $c_{i8}$, $i=1,2$ or even the function $f_2$.

In order to reduce again $\textbf{V}_f$ to an operator $\textbf{V}_{f_3}$ as simple as possible and arrive at the third basis vector $\textbf{Y}_3$ the inner automorphism (\ref{autof}) is applied, leaving $\textbf{Y}_1$ and $\textbf{Y}_2$ unchanged. For $f_3=\widetilde{f}$, this means a function $g$ is required that satisfies the following system of first-order linear PDEs
\begin{align}
\label{con3d2}
\begin{split}
U_1g&=n_1\\
U_2g&=n_2\\
U_3g&=n_3+f-f_3
\end{split}
\end{align}
where $n_1$, $n_2$ and $n_3$ are arbitrary constants, $f$ is the general solution to (\ref{con3d1}) and $f_3$ can be any solution to (\ref{con3d1}). Note that at this stage the structure constants $\alpha_i,\beta_i,\gamma_i$, $i=1,2$, the reduced parameters $c_{ij}$, $j=1,\ldots,9$ that define $U_i$, $i=1,2,3$ as well as the reduced function $f_2$ are all given. Now the compatibility of the first two equations of (\ref{con3d2}) is guaranteed if $\widetilde{m}=0$, where $\widetilde{m}=n_1\alpha+n_2\beta+\epsilon_{ij3}n_i(c_{j8}-2c_{j7})$. The rest compatibility conditions of system (\ref{con3d2}) yield another two equations, namely $(U_i-\gamma_i)(f-f_3)=\widetilde{l}_i$, $i=1,2$, where $\widetilde{l}_i$ are always constant expressed as $\widetilde{l}_i=n_1\alpha_i+n_2\beta_i+n_3\gamma_i+\epsilon_{ij3}\epsilon_{jkl}n_l(c_{k8}-2c_{k7})$, $i=1,2$. Consequently, quite similar to the previous subsection the general solution of the homogeneous counterpart of (\ref{con3d1}), i.e. for $l_1=l_2=f_2=0$, can always be removed from $f$. Therefore, the function $f_3$ of $\textbf{Y}_3$ can be chosen either as\,: $i)$ the trivial solution of the homogeneous counterpart of (\ref{con2d1}) when $\beta_1f_2=\beta_2f_2=U_3f_2=0$ and the substitutions $\widetilde{l}_1=l_1$ and $\widetilde{l}_2=l_2$ can be made, or $ii)$ a particular solution of (\ref{con3d1}) otherwise.

As the pairs of the operators $\textbf{Y}_1$ and $\textbf{Y}_2$ run Table \ref{optimal2d}, all three-dimensional equivalence algebras that originate from two-dimensional ones are obtained.

To this list, as previously mentioned, we must add $\mathfrak{so}(3)$ generated here by the rotations $\textbf{V}_4$, $\textbf{V}_5$ and $\textbf{V}_6$ that does not contain any two-dimensional subalgebras. The structure of this algebra allows no linear combination of any of these three operators with any of the rest equivalence generators in the finite-dimensional algebra $\bar{\mathfrak{g}}$. For the treatment of the infinite-dimensional one, since according to subsection \ref{op1d} it can always be eliminated from one of the operators, consider $\textbf{V}_4$, $\textbf{V}_5+\textbf{V}_f$ and $\textbf{V}_6+\textbf{V}_h$. In this case, the commutation relations taken together result in a system of three first-order linear PDEs for both $f$ and $h$, $U_1f-h=l_1$, $U_1h+f=l_2$ and $U_2h-U_3f=l_3$, where $l_i$ are arbitrary constants and $U_i=\textbf{V}_{i+3}$. And in order to reduce simultaneously $f$ and $h$ under (\ref{autof}) to $f_2=\widetilde{f}$ and $f_3=\widetilde{h}$, respectively, a function $g$ must satisfy the equations $U_1g=n_1$, $U_2g=n_2+f-f_2$ and $U_3g=n_3+h-f_3$ in place of system (\ref{con3d2}). Investigation of this system and its compatibility conditions, $U_1(f-f_2)-(h-f_3)=n_3$, $U_1(h-f_3)+(f-f_2)=-n_2$ and $U_2(h-f_3)-U_3(f-f_2)=n_1$, shows that for $n_1=l_3$, $n_2=-l_2$ and $n_3=l_1$ the gauge functions can be reduced down to $f_2=\lambda yr\rho^{-2}$ and $f_3=\lambda xr\rho^{-2}$, where $r=\sqrt{\rho^2+z^2}$ and $\lambda$ is an arbitrary constant.

\vspace{0.5cm}

\renewcommand\arraystretch{1.3}
\begin{table}[ht]
\centering
\begin{tabular}{[c!l]}
\thickhline
& \multicolumn{1}{c]}{Operators} \\
\thickhline
~1~ & $\textbf{V}_3$, \qquad\qquad\qquad\qquad\qquad $\textbf{V}_4+k_1\textbf{V}_8$, \qquad $\textbf{V}_7+k_2\textbf{V}_8$, ~\quad $k_1\neq0$ or $k_2\neq1/2,1,2$\\
\hline
2 & $\textbf{V}_3+\lambda\textbf{V}_9$, ~~\,\qquad\qquad\qquad $\textbf{V}_4$, ~\,\qquad\qquad\quad $2\textbf{V}_7+\textbf{V}_8$ \\
\hline
3 & $\textbf{V}_3+\lambda_1\textbf{V}_\phi-\lambda_2\textbf{V}_{\ln z}$, \qquad $\textbf{V}_4+\lambda_3\textbf{V}_9$, \qquad $\textbf{V}_7+\textbf{V}_8+\lambda_4\textbf{V}_9$ \\
\hline
4 & $\textbf{V}_3$, \qquad\qquad\qquad\qquad\qquad $\textbf{V}_4$, ~\,\qquad\qquad\quad $\textbf{V}_7+2\textbf{V}_8+\lambda\textbf{V}_\phi$ \\
\hline
5 & $\textbf{V}_4$, \qquad $\textbf{V}_5+\lambda\textbf{V}_{yr\rho^{-2}}$, \qquad $\textbf{V}_6+\lambda\textbf{V}_{xr\rho^{-2}}$\\
\hline
6 & $\textbf{V}_1$, \qquad $\textbf{V}_2$, \qquad\qquad\quad~ $\textbf{V}_4+k_1\textbf{V}_7+k_2\textbf{V}_8$, \,\,\qquad\qquad\qquad\qquad $k_1k_2\neq0, k_1\neq k_2$\\
\hline
7 & $\textbf{V}_1$, \qquad $\textbf{V}_2+\lambda\textbf{V}_x$, \,\,\,\qquad $\textbf{V}_4+k\textbf{V}_7+\lambda\textbf{V}_{(x^2-y^2)/2}$, \qquad\qquad\qquad $k\neq0$\\
\hline
8 & $\textbf{V}_1$, \qquad $\textbf{V}_2$, \qquad\qquad\quad~ $\textbf{V}_4+k\left(\textbf{V}_7+\textbf{V}_8+\lambda\textbf{V}_9\right)$, ~~\qquad\qquad\quad $k\neq0$\\
\hline
9 & $\textbf{V}_1$, \qquad $\textbf{V}_2$, \qquad\qquad\quad~ $\textbf{V}_4+k_1\textbf{V}_3+k_2\textbf{V}_8$, ~\qquad\qquad\qquad\qquad $k_1k_2\neq0$\\
\hline
10 & $\textbf{V}_1$, \qquad $\textbf{V}_2+\lambda_1\textbf{V}_x$, \qquad $\textbf{V}_4+k\textbf{V}_3+\lambda_2\textbf{V}_9+\lambda_1\textbf{V}_{(x^2-y^2)/2}$, \qquad $k\neq0$\\
\hline
11 & $\textbf{V}_2$, \quad\qquad\qquad~ $\textbf{V}_3$, \quad\qquad\qquad\, $\textbf{V}_7+k\textbf{V}_8$, ~~\qquad\qquad\qquad\qquad $k\neq0,1/2,1$\\
\hline
12 & $\textbf{V}_2+\lambda\textbf{V}_z$, ~\,\qquad $\textbf{V}_3$, \quad\qquad\qquad\, $\textbf{V}_7$ \\
\hline
13 & $\textbf{V}_2+\lambda_1\textbf{V}_9$, ~~\,\,\,\quad $\textbf{V}_3+\lambda_2\textbf{V}_9$, \qquad $2\textbf{V}_7+\textbf{V}_8$ \\
\hline
14 & $\textbf{V}_2-\lambda_2\textbf{V}_{\ln y}$, ~\quad $\textbf{V}_3-\lambda_3\textbf{V}_{\ln z}$, \,\quad $\textbf{V}_7+\textbf{V}_8+\lambda_1\textbf{V}_9$ \\
\hline
15 & $\textbf{V}_1+k_1\textbf{V}_8$, \,\qquad\qquad\qquad\qquad $\textbf{V}_2+k_2\textbf{V}_8$, \qquad\qquad~ $\textbf{V}_3+k_3\textbf{V}_8$, \qquad $k_3\neq0$\\
\hline
16 & $\textbf{V}_1+\lambda_1\textbf{V}_9+\lambda_4\textbf{V}_y+\lambda_5\textbf{V}_z$, \quad $\textbf{V}_2+\lambda_2\textbf{V}_9+\lambda_6\textbf{V}_z$, \quad $\textbf{V}_3+\lambda_3\textbf{V}_9$ \\
\thickhline
\end{tabular}
\caption{Optimal system of three-dimensional equivalence subalgebras.}
\label{optimal3d}
\end{table}
\renewcommand\arraystretch{1.0}



In conclusion, the optimal system of three-dimensional equivalence subalgebras comprises of the algebras listed in the Table \ref{optimal3d}, where $k$'s and $\lambda$'s are constants. The new contribution of the infinite-dimensional algebra to the third basis vector can be seen in cases 5, 7, 10, 14 and 16 expressed through the operators $\textbf{V}_{yr\rho^{-2}}=\rho^{-2}\left[xys\partial_{A^1}+(r+sy^2)\partial_{A^2}+yzr^{-1}\partial_{A^3}\right]$, $\textbf{V}_{xr\rho^{-2}}=\rho^{-2}\left[(r+sx^2)\partial_{A^1}+xys\partial_{A^2}+xzr^{-1}\partial_{A^3}\right]$, $\textbf{V}_{(x^2-y^2)/2}=x\partial_{A^1}-y\partial_{A^2}$ and similar to the prior ones, $\textbf{V}_{\ln y}=y^{-1}\partial_{A^2}$, $\textbf{V}_x=\partial_{A^1}$ and $\textbf{V}_y=\partial_{A^2}$, where $s=r^{-1}-2r\rho^{-2}$.

\begin{remark}\normalfont
From the classifying equations (\ref{APheq}), the values $k_1=0$ and $k=0$ in cases 9 and 10, respectively, of Table \ref{optimal3d} yield zero vector potentials for the corresponding symmetry groups and are thus excluded.
\end{remark}


\section{Symmetry Classification}\label{symmetryclasses}

Once the classification of the equivalence algebra is made, now we return to section \ref{classifying} and, in particular, the classifying equations (\ref{APheq}). As explained there, the projection of the equivalence subalgebras spanned by any $\textbf{V}$ of the form (\ref{eq}) yield symmetry subalgebras spanned by $\textbf{v}$ (\ref{sym}) if and only if the potentials $\boldsymbol{A}$ and $\Phi$ satisfy the corresponding subsystem of (\ref{APheq}). The symmetry algebras obtained are extensions of the principal Lie algebra consisting only of the generator $\textbf{v}_0$ of time translations.

\subsection{Systems with two Lie point symmetries}

Consider the symmetry generators $\textbf{X}$ coming from the optimal system of one-dimensional subalgebras of Table \ref{optimal1d}, which yield a second symmetry for system (\ref{primpotsystemeq}) besides time translations. Solutions of equations (\ref{APheq}) in this case can be recovered from \cite{me} (see Section 3 therein). The results are shown in the next table, where $F_i=F_i(u_1,u_2)$ and $G=G(u_1,u_2)$ are arbitrary functions, while $\boldsymbol{F}=(F_1,F_2,F_3)$ is an arbitrary vector function of the same form. For brevity, we also use cylindrical coordinates $(\rho,\phi,z)$ in many cases. Finally $k$'s and $\lambda$'s are constants, the difference being that the former also define the symmetry.

\vspace{0.4cm}
 
\renewcommand{\arrayrulewidth}{1pt}
\renewcommand\arraystretch{1.15}
\begin{longtable}{|c|l|l|l|}
\hline
&\multicolumn{1}{c|}{Symmetry generator} & \multicolumn{2}{c|}{Electromagnetic potential}  \\
\hline
~1~ & \specialcell{$\textbf{X}=\textbf{v}_4+k_1\textbf{v}_7+k_2\textbf{v}_8$ \\ $k_2\neq k_1\neq0$} & \specialcell{$A^1=z^{-\frac{k_2}{k_1}}\left(xF_1-yF_2\right)$ \\ $A^2=z^{-\frac{k_2}{k_1}}\left(yF_1+xF_2\right)$ \\ $A^3=z^{1-\frac{k_2}{k_1}}F_3$ \\ $~\,\Phi=z^{2\left(1-\frac{k_2}{k_1}\right)}G$} & \specialcell{$u_1=\rho/z$ \\ $u_2=\ln z-k_1\phi$}  \\
\hline
2 & \specialcell{$\textbf{X}=\textbf{v}_4+k\left(\textbf{v}_7+\textbf{v}_8\right)$ \\ $k\neq0$} & \specialcell{$A^1=z^{-1}\left(xF_1-yF_2\right)$ \\ $A^2=z^{-1}\left(yF_1+xF_2\right)$ \\ $A^3=F_3$ \\ $~\,\Phi=\lambda\ln z+G$} & \specialcell{$u_1=\rho/z$ \\ $u_2=\ln z-k\phi$}  \\
\hline
3 & \specialcell{$\textbf{X}=\textbf{v}_4+k_1\textbf{v}_3+k_2\textbf{v}_8$ \\ $k_2\neq0$} & \specialcell{$A^1=e^{-k_2\phi}\left(xF_1-yF_2\right)$ \\ $A^2=e^{-k_2\phi}\left(yF_1+xF_2\right)$ \\ $A^3=e^{-k_2\phi}F_3$ \\ $~\,\Phi=e^{-2k_2\phi}G$} & \specialcell{$u_1=\rho$ \\ $u_2=z-k_1\phi$}    \\
\hline
4 &$\textbf{X}=\textbf{v}_4+k\textbf{v}_3$ & \specialcell{$A^1=xF_1-yF_2$ \\ $A^2=yF_1+xF_2$ \\ $A^3=F_3$ \\ $~\,\Phi=\lambda\phi+G$} & \specialcell{$u_1=\rho$ \\ $u_2=z-k\phi$}   \\
\hline
5 &$\textbf{X}=\textbf{v}_7+k\textbf{v}_8$, \quad$k\neq1$ & \specialcell{$~\boldsymbol{A}=x^{1-k}\boldsymbol{F}$ \\ $~\,\Phi=x^{2(1-k)}G$} & \specialcell{$u_1=y/x$ \\ $u_2=z/y$}  \\
\hline
6 &$\textbf{X}=\textbf{v}_7+\textbf{v}_8$ & \specialcell{$~\boldsymbol{A}=\boldsymbol{F}$ \\ $~\,\Phi=\lambda\ln z+G$} & \specialcell{$u_1=y/x$ \\ $u_2=z/y$}  \\
\hline
7 &$\textbf{X}=\textbf{v}_3+k\textbf{v}_8$, \quad$k\neq0$ & \specialcell{$~\boldsymbol{A}=e^{-kz}\boldsymbol{F}$ \\ $~\,\Phi=e^{-2kz}G$} & \specialcell{$u_1=x$ \\ $u_2=y$} \\
\hline
8 &$\textbf{X}=\textbf{v}_3$ & \specialcell{$~\boldsymbol{A}=\boldsymbol{F}$ \\ $~\,\Phi=\lambda z+G$} & \specialcell{$u_1=x$ \\ $u_2=y$}\\
\hline
\caption{Vector and scalar potentials of the electromagnetic field for two-parameter symmetry groups generated by $\textbf{v}_0$ and $\textbf{X}$.}
\label{sym2d}
\end{longtable}
\renewcommand\arraystretch{1.0}

\subsection{Systems with three Lie point symmetries}

Here, we consider the symmetry algebras spanned by $\textbf{X}_1$ and $\textbf{X}_2$ coming from the optimal system of two-dimensional subalgebras of Table \ref{optimal2d}, which give two more symmetries for system (\ref{primpotsystemeq}) besides time translations. In this case, taking one of the equivalence generators we arrive at a solution of equations (\ref{APheq}), that is, one of the potentials listed in Table \ref{sym2d}. Then each of these solutions is replaced back to the subsystem of (\ref{APheq}) corresponding to the second generator. Solutions to the latter equations are rather easily found usually by eliminating one of the variables $u_1$ or $u_2$ of the potentials. The results are shown in the next table, using the same notation with the previous table only now $F_i=F_i(u)$ and $G=G(u)$.

\vspace{0.4cm}

\renewcommand\arraystretch{1.01}
\begin{longtable}{|c|l|l|l|}
\hline
&\multicolumn{1}{c|}{Symmetry generators} & \multicolumn{2}{c|}{Electromagnetic potential}  \\
\hline
~1~ &\specialcell{$\textbf{X}_1=\textbf{v}_3$ \\ $\textbf{X}_2=\textbf{v}_4+k_1\textbf{v}_7+k_2\textbf{v}_8$ \\ $2k_2,k_2\neq k_1\neq0$}& \specialcell{$A^1=e^{-k_2\phi}\left(xF_1-yF_2\right)$ \\ $A^2=e^{-k_2\phi}\left(yF_1+xF_2\right)$ \\ $A^3=e^{\left(k_1-k_2\right)\phi}F_3$ \\ $~\,\Phi=e^{2\left(k_1-k_2\right)\phi}G$} & $u=\ln\rho-k_1\phi$   \\
\hline
2 &\specialcell{$\textbf{X}_1=\textbf{v}_3$ \\ $\textbf{X}_2=\textbf{v}_4+k\left(2\textbf{v}_7+\textbf{v}_8\right)$ \\ $k\neq0$}  & \specialcell{$A^1=e^{-k\phi}\left(xF_1-yF_2\right)$ \\ $A^2=e^{-k\phi}\left(yF_1+xF_2\right)$ \\ $A^3=e^{k\phi}F_3$                          \\ $~\,\Phi=\lambda z+e^{2k\phi}G$}        & $u=\ln\rho-2k\phi$   \\
\hline
3 &\specialcell{$\textbf{X}_1=\textbf{v}_3$ \\ $\textbf{X}_2=\textbf{v}_4+k\left(\textbf{v}_7+\textbf{v}_8\right)$ \\ $k\neq0$}  & \specialcell{$A^1=\rho^{-1}\left[xF_1-y\left(F_2+\lambda_1z\rho^{-1}\right)\right]$   \\ $A^2=\rho^{-1}\left[yF_1+x\left(F_2+\lambda_1z\rho^{-1}\right)\right]$   \\ $A^3=F_3$   \\ $~\,\Phi=\lambda_2\ln\rho+G$} & $u=\ln\rho-k\phi$ \\
\hline
4 &\specialcell{$\textbf{X}_1=\textbf{v}_3+k_1\textbf{v}_8$ \\ $\textbf{X}_2=\textbf{v}_4+k_2\textbf{v}_8$ \\ $k_1\neq0\text{ or }k_2\neq0$} & \specialcell{$A^1=e^{-\left(k_1z+k_2\phi\right)}\left(xF_1-yF_2\right)$ \\ $A^2=e^{-\left(k_1z+k_2\phi\right)}\left(yF_1+xF_2\right)$ \\ $A^3=e^{-\left(k_1z+k_2\phi\right)}F_3$ \\ $~\,\Phi=e^{-2\left(k_1z+k_2\phi\right)}G$} & $u=\rho$    \\
\hline
5 &\specialcell{$\textbf{X}_1=\textbf{v}_3$ \\ $\textbf{X}_2=\textbf{v}_4$} & \specialcell{$A^1=xF_1-yF_2-\lambda_3yz\rho^{-2}$ \\ $A^2=yF_1+xF_2+\lambda_3yz\rho^{-2}$ \\ $A^3=F_3$ \\ $~\,\Phi=\lambda_1z+\lambda_2\phi+G$} & $u=\rho$    \\
\hline
6 &\specialcell{$\textbf{X}_1=\textbf{v}_4+k_1\textbf{v}_8$ \\ $\textbf{X}_2=\textbf{v}_7+k_2\textbf{v}_8$ \\ $k_1\neq0\text{ or }k_2\neq1,2$} & \specialcell{$A^1=z^{-k_2}e^{-k_1\phi}\left(xF_1-yF_2\right)$ \\ $A^2=z^{-k_2}e^{-k_1\phi}\left(yF_1+xF_2\right)$ \\ $A^3=z^{1-k_2}e^{-k_1\phi}F_3$ \\ $~\,\Phi=z^{2(1-k_2)}e^{-2k_1\phi}G$} & $u=\rho/z$   \\
\hline
7 &\specialcell{$\textbf{X}_1=\textbf{v}_4$ \\ $\textbf{X}_2=\textbf{v}_7+\textbf{v}_8$} & \specialcell{$A^1=z^{-1}\left(xF_1-yF_2\right)$ \\ $A^2=z^{-1}\left(yF_1+xF_2\right)$ \\ $A^3=F_3$ \\ $~\,\Phi=\lambda_1\phi+\lambda_2\ln z+G$} & $u=\rho/z$   \\
\hline
8 &\specialcell{$\textbf{X}_1=\textbf{v}_4$ \\ $\textbf{X}_2=\textbf{v}_7+2\textbf{v}_8$} & \specialcell{$A^1=\rho^{-2}\left[xF_1-y\left(F_2+\lambda\ln\rho\right)\right]$ \\ $A^2=\rho^{-2}\left[yF_1+x\left(F_2+\lambda\ln\rho\right)\right]$ \\ $A^3=z^{-1}F_3$ \\ $~\,\Phi=z^{-2}G$} & $u=\rho/z$   \\
\hline
9 &\specialcell{$\textbf{X}_1=\textbf{v}_3$ \\ $\textbf{X}_2=\textbf{v}_7+k\textbf{v}_8$ \\ $k\neq1,1/2$} &  \specialcell{$~\boldsymbol{A}=x^{1-k}\boldsymbol{F}$ \\ $~\,\Phi=x^{2(1-k)}G$} & $u=y/x$ \\
\hline
10 &\specialcell{$\textbf{X}_1=\textbf{v}_3$ \\ $\textbf{X}_2=\textbf{v}_7+\textbf{v}_8$} & \specialcell{$A^1=F_1$ \\ $A^2=F_2$ \\ $A^3=\lambda_1\ln y+F_3$ \\ $~\,\Phi=\lambda_2\ln y+G$} & $u=y/x$ \\
\hline
11 &\specialcell{$\textbf{X}_1=\textbf{v}_3$ \\ $\textbf{X}_2=2\textbf{v}_7+\textbf{v}_8$} & \specialcell{$~\boldsymbol{A}=\sqrt{x}\boldsymbol{F}$ \\ $~\,\Phi=\lambda z+xG$} & $u=y/x$ \\
\hline
12 &\specialcell{$\textbf{X}_1=\textbf{v}_2+k_1\textbf{v}_8$ \\ $\textbf{X}_2=\textbf{v}_3+k_2\textbf{v}_8$, \, $k_2\neq0$} & \specialcell{$~\boldsymbol{A}=e^{-\left(k_1y+k_2z\right)}\boldsymbol{F}$ \\ $~\,\Phi=e^{-2\left(k_1y+k_2z\right)}G$} & $u=x$ \\
\hline
13 &\specialcell{$\textbf{X}_1=\textbf{v}_2$ \\ $\textbf{X}_2=\textbf{v}_3$} & \specialcell{$A^1=0$ \\ $A^2=F_2$ \\ $A^3=\lambda_3y+F_3$ \\ $~\,\Phi=\lambda_1y+\lambda_2z+G$} & $u=x$ \\
\hline
\caption{Vector and scalar potentials of the electromagnetic field for three-parameter symmetry groups generated by $\textbf{v}_0$, $\textbf{X}_1$ and $\textbf{X}_2$.}  
\label{sym3d}
\end{longtable}
\renewcommand\arraystretch{1.0}

\begin{remark}\normalfont
Note that the sixth case of the above table for $k_1=0$, $k_2=3$, $F_1=F_3=0$ and $F_2(u)=(u^2+1)^{-3/2}$ recovers the vector potential of the magnetic dipole, $\boldsymbol{A}=\left(-\,y,x,0\right)/r^3$. This corresponds to the so-called St\"ormer problem which admits the symmetry generators $\textbf{v}_0$, $\textbf{X}_1=\textbf{v}_4$ and $\textbf{X}_2=\textbf{v}_7+3\textbf{v}_8$. And as we will see in the next section, although the first two generate Noether symmetries and correspond to the Hamiltonian $H$ and the well-known integral of motion $I=x\dot{y}-y\dot{x}+\rho^2/r^3$, respectively, the latter does not.
\end{remark}

\subsection{Systems with four Lie point symmetries}

Finally, consider the symmetry algebras spanned by $\textbf{X}_1$, $\textbf{X}_2$ and $\textbf{X}_3$ that originate from the optimal system of three-dimensional subalgebras of Table \ref{optimal3d}, which give three more symmetries for system (\ref{primpotsystemeq}) besides time translations. Now each of the potentials given in the previous subsection is replaced back to the subsystem of (\ref{APheq}) for the third generator. Solutions to the latter equations yield the results shown in the next table, where $a_i$, $i=1,\ldots,4$ are arbitrary constants. Some classes may of course admit more symmetries, when they no longer correspond to magnetic fields which are either curved or inhomogeneous. For example, cases 11 and 13-15 of the table below amount to straight magnetic fields, while cases 12 and 16 result in linear equations of motion.

\vspace{0.4cm}

\renewcommand\arraystretch{1.11}
\begin{longtable}{|c|l|l|}
\hline
&\multicolumn{1}{c|}{Symmetry generators} & \multicolumn{1}{c|}{Electromagnetic potential}  \\
\hline
~1~ &\specialcell{$\textbf{X}_1=\textbf{v}_3$ \\ $\textbf{X}_2=\textbf{v}_4+k_1\textbf{v}_8$ \\ $\textbf{X}_3=\textbf{v}_7+k_2\textbf{v}_8$ \\ $k_1\neq0$ or $k_2\neq1/2,1,2$}& \specialcell{$A^1=e^{-k_1\phi}\rho^{-k_2}\left(a_1x-a_2y\right)$ \\ $A^2=e^{-k_1\phi}\rho^{-k_2}\left(a_1y+a_2x\right)$ \\ $A^3=a_3e^{-k_1\phi}\rho^{1-k_2}$ \\ $~\,\Phi=a_4e^{-2k_1\phi}\rho^{2(1-k_2)}$}   \\
\hline
2 &\specialcell{$\textbf{X}_1=\textbf{v}_3$ \\ $\textbf{X}_2=\textbf{v}_4$ \\ $\textbf{X}_3=2\textbf{v}_7+\textbf{v}_8$ }& \specialcell{$A^1=\sqrt{\rho}^{-1}\left(a_1x-a_2y\right)$ \\ $A^2=\sqrt{\rho}^{-1}\left(a_1y+a_2x\right)$ \\ $A^3=a_3\sqrt{\rho}$ \\ $~\,\Phi=\lambda z+a_4\rho$}   \\
\hline
3 &\specialcell{$\textbf{X}_1=\textbf{v}_3$ \\ $\textbf{X}_2=\textbf{v}_4$ \\ $\textbf{X}_3=\textbf{v}_7+\textbf{v}_8$ }& \specialcell{$A^1=\rho^{-1}\left[a_1x-y\left(a_2+\lambda_1 z\rho^{-1}\right)\right]$ \\ $A^2=\rho^{-1}\left[a_1y+x\left(a_2+\lambda_1 z\rho^{-1}\right)\right]$ \\ $A^3=\lambda_2\ln\rho$ \\ $~\,\Phi=\lambda_3\phi+\lambda_4\ln\rho$}   \\
\hline
4 &\specialcell{$\textbf{X}_1=\textbf{v}_3$ \\ $\textbf{X}_2=\textbf{v}_4$ \\ $\textbf{X}_3=\textbf{v}_7+2\textbf{v}_8$ }& \specialcell{$A^1=\rho^{-2}\left[a_1x-y\left(a_2+\lambda\ln\rho\right)\right]$ \\ $A^2=\rho^{-2}\left[a_1y+x\left(a_2+\lambda\ln\rho\right)\right]$ \\ $A^3=a_3\rho^{-1}$ \\ $~\,\Phi=a_4\rho^{-2}$}   \\
\hline
5 &\specialcell{$\textbf{X}_1=\textbf{v}_4$ \\ $\textbf{X}_2=\textbf{v}_5$ \\ $\textbf{X}_3=\textbf{v}_6$ }& \specialcell{$A^1=~~\,\lambda yzr^{-1}\rho^{-2}$ \\ $A^2=-\lambda xzr^{-1}\rho^{-2}$ \\ $A^3=0$ \\ $~\,\Phi=G(r)$}   \\
\hline
6 &\specialcell{$\textbf{X}_1=\textbf{v}_1$ \\ $\textbf{X}_2=\textbf{v}_2$ \\ $\textbf{X}_3=\textbf{v}_4+k_1\textbf{v}_7+k_2\textbf{v}_8$ \\ $k_1k_2\neq0, k_1\neq k_2$ }& \specialcell{$A^1=a_1z^{1-\frac{k_2}{k_1}}\cos\left(\ln(z/k_1)+a_2\right)$ \\ $A^2=a_1z^{1-\frac{k_2}{k_1}}\sin\left(\ln(z/k_1)+a_2\right)$ \\ $A^3=0$ \\ $~\,\Phi=a_4z^{2\left(1-\frac{k_2}{k_1}\right)}$}   \\
\hline
7 &\specialcell{$\textbf{X}_1=\textbf{v}_1$ \\ $\textbf{X}_2=\textbf{v}_2$ \\ $\textbf{X}_3=\textbf{v}_4+k\textbf{v}_7$ \\ $k\neq0$ }& \specialcell{$A^1=a_1z\cos\left(\ln(z/k)+a_2\right)+\lambda y$ \\ $A^2=a_1z\sin\left(\ln(z/k)+a_2\right)$ \\ $A^3=0$ \\ $~\,\Phi=a_4z^2$}   \\
\hline
8 &\specialcell{$\textbf{X}_1=\textbf{v}_1$ \\ $\textbf{X}_2=\textbf{v}_2$ \\ $\textbf{X}_3=\textbf{v}_4+k\left(\textbf{v}_7+\textbf{v}_8\right)$ \\ $k\neq0$ }& \specialcell{$A^1=a_1\cos\left(\ln(z/k)+a_2\right)$ \\ $A^2=a_1\sin\left(\ln(z/k)+a_2\right)$ \\ $A^3=0$ \\ $~\,\Phi=\lambda\ln z$}  \\
\hline
9 &\specialcell{$\textbf{X}_1=\textbf{v}_1$ \\ $\textbf{X}_2=\textbf{v}_2$ \\ $\textbf{X}_3=\textbf{v}_4+k_1\textbf{v}_3+k_2\textbf{v}_8$ \\ $k_1k_2\neq0$ }& \specialcell{$A^1=a_1e^{-\frac{k_2}{k_1}z}\cos\left(z/k_1+a_2\right)$ \\ $A^2=a_1e^{-\frac{k_2}{k_1}z}\sin\left(z/k_1+a_2\right)$ \\ $A^3=0$ \\ $~\,\Phi=a_4e^{-2\frac{k_2}{k_1}z}$}   \\
\hline
10 &\specialcell{$\textbf{X}_1=\textbf{v}_1$ \\ $\textbf{X}_2=\textbf{v}_2$ \\ $\textbf{X}_3=\textbf{v}_4+k\textbf{v}_3$ \\ $k\neq0$ }& \specialcell{$A^1=a_1\cos\left(z/k+a_2\right)+\lambda_1y$ \\ $A^2=a_1\sin\left(z/k+a_2\right)$ \\ $A^3=0$ \\ $~\,\Phi=\lambda_2z/k$}  \\
\hline
11 &\specialcell{$\textbf{X}_1=\textbf{v}_2$ \\ $\textbf{X}_2=\textbf{v}_3$ \\ $\textbf{X}_3=\textbf{v}_7+k\textbf{v}_8$ \\ $k\neq0,1/2,1$ }& \specialcell{$A^1=0$ \\ $A^2=a_2x^{1-k}$ \\ $A^3=a_3x^{1-k}$ \\ $~\,\Phi=a_4x^{2(1-k)}$} \\
\hline
12 &\specialcell{$\textbf{X}_1=\textbf{v}_2$ \\ $\textbf{X}_2=\textbf{v}_3$ \\ $\textbf{X}_3=\textbf{v}_7$}& \specialcell{$A^1=0$ \\ $A^2=a_2x$ \\ $A^3=a_3x+\lambda y$ \\ $~\,\Phi=a_4x^2$} \\
\hline
13 &\specialcell{$\textbf{X}_1=\textbf{v}_2$ \\ $\textbf{X}_2=\textbf{v}_3$ \\ $\textbf{X}_3=2\textbf{v}_7+\textbf{v}_8$}& \specialcell{$A^1=0$ \\ $A^2=a_2\sqrt{x}$ \\  $A^3=a_3\sqrt{x}$ \\ $~\,\Phi=a_4x+\lambda_1y+\lambda_2z$}  \\
\hline
14 &\specialcell{$\textbf{X}_1=\textbf{v}_2$ \\ $\textbf{X}_2=\textbf{v}_3$ \\ $\textbf{X}_3=\textbf{v}_7+\textbf{v}_8$}& \specialcell{$A^1=0$ \\ $A^2=\lambda_2\ln x$ \\  $A^3=\lambda_3\ln x$ \\ $~\,\Phi=\lambda_1\ln x$} \\
\hline
15 &\specialcell{$\textbf{X}_1=\textbf{v}_1+k_1\textbf{v}_8$ \\ $\textbf{X}_2=\textbf{v}_2+k_2\textbf{v}_8$ \\ $\textbf{X}_3=\textbf{v}_3+k_3\textbf{v}_8$ \\ $k_3\neq0$}& \specialcell{$A^1=a_1e^{-(k_1x+k_2y+k_3z)}$ \\ $A^2=a_2e^{-(k_1x+k_2y+k_3z)}$ \\  $A^3=a_3e^{-(k_1x+k_2y+k_3z)}$ \\ $~\,\Phi=a_4e^{-2(k_1x+k_2y+k_3z)}$}  \\
\hline
16 &\specialcell{$\textbf{X}_1=\textbf{v}_1$ \\ $\textbf{X}_2=\textbf{v}_2$ \\ $\textbf{X}_3=\textbf{v}_3$ }& \specialcell{$A^1=0$ \\ $A^2=\lambda_4x$ \\  $A^3=\lambda_5x+\lambda_6y$ \\ $~\,\Phi=\lambda_1x+\lambda_2y+\lambda_3z$}  \\
\hline
\caption{Vector and scalar potentials of the electromagnetic field for four-parameter symmetry groups generated by $\textbf{v}_0$, $\textbf{X}_1$, $\textbf{X}_2$ and $\textbf{X}_3$.}  
\label{sym4d}
\end{longtable}
\renewcommand\arraystretch{1.0}

\section{Classification in terms of Noether symmetries}\label{noetherclasses}

In light of \cite{me}, a Noether symmetry classification can also be given and rather easily too. Since, as demonstrated there, the Noether symmetry condition required only the extra constraint $c_8=2c_7$. From the latter and the previous three tables we can thus classify system (\ref{primpotsystemeq}) in terms of Noether symmetries. Following the same notation with the previous section, the results are collected in the next three tables.

\vspace{0.3cm}
 
\renewcommand{\arrayrulewidth}{1pt}
\renewcommand\arraystretch{1.15}
\begin{longtable}{|c|l|l|l|} 
\hline
&\multicolumn{1}{c|}{Noether symmetry} & \multicolumn{2}{c|}{Electromagnetic potential}  \\
\hline
~1~ &\specialcell{$\textbf{X}=\textbf{v}_4+k\left(\textbf{v}_7+2\textbf{v}_8\right)$ \\ $k\neq0$} & \specialcell{$A^1=z^{-2}\left(xF_1-yF_2\right)$ \\ $A^2=z^{-2}\left(yF_1+xF_2\right)$ \\ $A^3=z^{-1}F_3$ \\ $~\,\Phi=z^{-2}G$} & \specialcell{$u_1=\rho/z$ \\ $u_2=\ln z-k\phi$}  \\
\hline
2 &$\textbf{X}=\textbf{v}_4+k\textbf{v}_3$ & \specialcell{$A^1=xF_1-yF_2$ \\ $A^2=yF_1+xF_2$ \\ $A^3=F_3$ \\ $~\,\Phi=\lambda\phi+G$} & \specialcell{$u_1=\rho$ \\ $u_2=z-k\phi$}   \\
\hline
3 &$\textbf{X}=\textbf{v}_7+2\textbf{v}_8$ & \specialcell{$~\boldsymbol{A}=x^{-1}\boldsymbol{F}$ \\ $~\,\Phi=x^{-2}G$} & \specialcell{$u_1=y/x$ \\ $u_2=z/y$}  \\
\hline
4 &$\textbf{X}=\textbf{v}_3$ & \specialcell{$~\boldsymbol{A}=\boldsymbol{F}$ \\ $~\,\Phi=\lambda z+G$} & \specialcell{$u_1=x$ \\ $u_2=y$}\\
\hline
\caption{Vector and scalar potentials of the electromagnetic field for two-parameter Noether symmetry groups generated by $\textbf{v}_0$ and $\textbf{X}$.}
\label{nsym2d}
\end{longtable}
\renewcommand\arraystretch{1.0}



For equations (\ref{primpotsystemeq}), the invariants $I$ that correspond to Noether point symmetries are given by the formula (see \cite{me})\vspace{-0.15cm}
\begin{equation}
\label{integral}
I(t,\boldsymbol{x},\dot{\boldsymbol{x}})=\left(c_7x^i-\epsilon_{ijk}c_{7-k}x^j+c_i\right)\left(\dot{x}^i+A^i(\boldsymbol{x})\right)-\left(2c_7t+c_0\right)H(\boldsymbol{x},\dot{\boldsymbol{x}})+c_9t-f(\boldsymbol{x})
\end{equation}
Here $H(\boldsymbol{x},\dot{\boldsymbol{x}})=\dot{\boldsymbol{x}}^2/2+\Phi(\boldsymbol{x})$ is the well-known Hamiltonian function that corresponds to the Noether symmetry generator $\textbf{v}_0$. We also recall that the Poisson bracket admitted by the system is
\begin{equation}
\label{poisson}
\left\{I_1,I_2\right\}=\frac{\partial I_1}{\partial x^i}\frac{\partial I_2}{\partial\dot{x}^i}-\frac{\partial I_1}{\partial\dot{x}^i}\frac{\partial I_2}{\partial x^i}+\epsilon_{ijk}B^k\frac{\partial I_1}{\partial\dot{x}^i}\frac{\partial I_2}{\partial\dot{x}^j}
\end{equation}
for any two functions $I_1(\boldsymbol{x},\dot{\boldsymbol{x}})$ and $I_2(\boldsymbol{x},\dot{\boldsymbol{x}})$. Thus, from (\ref{integral}), (\ref{poisson}) and the previous table, the following conclusion is drawn regarding the existence of an additional constant of motion besides the energy $H$ of the system.


\begin{corollary}\normalfont
For inhomogeneous and curved magnetic fields, the autonomous system (\ref{primpotsystemeq}) of charged particle motion admits a first integral of motion $I$ that corresponds to a Noether point symmetry $\textbf{X}$, which is functionally independent of the Hamiltonian function $H$ and in involution with it, in two representative cases under the equivalence transformations (\ref{group})\,:
\begin{enumerate}
	\item case 2 of Table \ref{nsym2d} for $\lambda=0$, where $I=x\dot{y}-y\dot{x}+k\dot{z}+(x^2+y^2)F_2+kF_3$,
	\item case 4 of Table \ref{nsym2d} for $\lambda=0$, where $I=\dot{z}+F_3$.
\end{enumerate}
\end{corollary}



\renewcommand\arraystretch{1.23}
\begin{longtable}{|c|l|l|l|}
\hline
&\multicolumn{1}{c|}{Noether symmetries} & \multicolumn{2}{c|}{Electromagnetic potential}  \\
\hline
~1~ &\specialcell{$\textbf{X}_1=\textbf{v}_3$ \\ $\textbf{X}_2=\textbf{v}_4+k\left(\textbf{v}_7+2\textbf{v}_8\right)$ \\ $k\neq0$} & \specialcell{$A^1=e^{-2k\phi}\left(xF_1-yF_2\right)$ \\ $A^2=e^{-2k\phi}\left(yF_1+xF_2\right)$ \\ $A^3=e^{-k\phi}F_3$ \\ $~\,\Phi=e^{-2k\phi}G$} & $u=\ln\rho-k\phi$   \\
\hline
2 &\specialcell{$\textbf{X}_1=\textbf{v}_3$ \\ $\textbf{X}_2=\textbf{v}_4$} & \specialcell{$A^1=xF_1-yF_2-\lambda_3yz\rho^{-2}$ \\ $A^2=yF_1+xF_2+\lambda_3xz\rho^{-2}$ \\ $A^3=F_3$ \\ $~\,\Phi=\lambda_1z+\lambda_2\phi+G$} & $u=\rho$    \\
\hline
3 &\specialcell{$\textbf{X}_1=\textbf{v}_4$ \\ $\textbf{X}_2=\textbf{v}_7+2\textbf{v}_8$} & \specialcell{$A^1=\rho^{-2}\left[xF_1-y\left(F_2+\lambda\ln\rho\right)\right]$ \\ $A^2=\rho^{-2}\left[yF_1+x\left(F_2+\lambda\ln\rho\right)\right]$ \\ $A^3=z^{-1}F_3$ \\ $~\,\Phi=z^{-2}G$} & $u=\rho/z$   \\
\hline
4 &\specialcell{$\textbf{X}_1=\textbf{v}_3$ \\ $\textbf{X}_2=\textbf{v}_7+2\textbf{v}_8$} &  \specialcell{$~\boldsymbol{A}=x^{-1}\boldsymbol{F}$ \\ $~\,\Phi=x^{-2}G$} & $u=y/x$ \\
\hline
5 &\specialcell{$\textbf{X}_1=\textbf{v}_2$ \\ $\textbf{X}_2=\textbf{v}_3$} & \specialcell{$A^1=0$ \\ $A^2=F_2$ \\ $A^3=\lambda_3y+F_3$ \\ $~\,\Phi=\lambda_1y+\lambda_2z+G$} & $u=x$ \\
\hline
\caption{Vector and scalar potentials of the electromagnetic field for three-parameter Noether symmetry groups generated by $\textbf{v}_0$, $\textbf{X}_1$ and $\textbf{X}_2$.}  
\label{nsym3d}
\end{longtable}
\renewcommand\arraystretch{1.0}

Subsequently, we can start investigating aspects of complete integrability in terms of Noether point symmetries. In other words, we study the construction of two additional first integrals of motion $I_1$ and $I_2$, which are functionally independent of the Hamiltonian $H$ and all three $H$, $I_1$ and $I_2$ are pairwise in involution, based on Noether point symmetries. First of all, the cases where the invariants $I_1$ and $I_2$ correspond directly to point symmetries $\textbf{X}_1$ and $\textbf{X}_2$, respectively, would lie among the potentials of Table \ref{nsym3d} for three-dimensional Noether symmetry algebras.

\vspace{0.4cm}

\renewcommand\arraystretch{1.25}
\begin{longtable}{|c|l|l|}
\hline
&\multicolumn{1}{c|}{Noether symmetries} & \multicolumn{1}{c|}{Electromagnetic potential}  \\
\hline
~1~ &\specialcell{$\textbf{X}_1=\textbf{v}_3$ \\ $\textbf{X}_2=\textbf{v}_4$ \\ $\textbf{X}_3=\textbf{v}_7+2\textbf{v}_8$ }& \specialcell{$A^1=\rho^{-2}\left[a_1x-y\left(a_2+\lambda\ln\rho\right)\right]$ \\ $A^2=\rho^{-2}\left[a_1y+x\left(a_2+\lambda\ln\rho\right)\right]$ \\ $A^3=a_3\rho^{-1}$ \\ $~\,\Phi=a_4\rho^{-2}$}   \\
\hline
2 &\specialcell{$\textbf{X}_1=\textbf{v}_4$ \\ $\textbf{X}_2=\textbf{v}_5$ \\ $\textbf{X}_3=\textbf{v}_6$ }& \specialcell{$A^1=~~\,\lambda yzr^{-1}\rho^{-2}$ \\ $A^2=-\lambda xzr^{-1}\rho^{-2}$ \\ $A^3=0$ \\ $~\,\Phi=G(r)$}   \\
\hline
3 &\specialcell{$\textbf{X}_1=\textbf{v}_1$ \\ $\textbf{X}_2=\textbf{v}_2$ \\ $\textbf{X}_3=\textbf{v}_4+k\left(\textbf{v}_7+2\textbf{v}_8\right)$ \\ $k\neq0$ }& \specialcell{$A^1=a_1z^{-1}\cos\left(\ln(z/k)+a_2\right)$ \\ $A^2=a_1z^{-1}\sin\left(\ln(z/k)+a_2\right)$ \\ $A^3=0$ \\ $~\,\Phi=a_4z^{-2}$}   \\
\hline
4 &\specialcell{$\textbf{X}_1=\textbf{v}_1$ \\ $\textbf{X}_2=\textbf{v}_2$ \\ $\textbf{X}_3=\textbf{v}_4+k\textbf{v}_3$ \\ $k\neq0$ }& \specialcell{$A^1=a_1\cos\left(z/k+a_2\right)+\lambda_1y$ \\ $A^2=a_1\sin\left(z/k+a_2\right)$ \\ $A^3=0$ \\ $~\,\Phi=\lambda_2z/k$}  \\
\hline
5 &\specialcell{$\textbf{X}_1=\textbf{v}_2$ \\ $\textbf{X}_2=\textbf{v}_3$ \\ $\textbf{X}_3=\textbf{v}_7+2\textbf{v}_8$ \\ $k\neq0,1/2,1$ }& \specialcell{$A^1=0$ \\ $A^2=a_2x^{-1}$ \\ $A^3=a_3x^{-1}$ \\ $~\,\Phi=a_4x^{-2}$} \\
\hline
6 &\specialcell{$\textbf{X}_1=\textbf{v}_1$ \\ $\textbf{X}_2=\textbf{v}_2$ \\ $\textbf{X}_3=\textbf{v}_3$ }& \specialcell{$A^1=0$ \\ $A^2=\lambda_4x$ \\  $A^3=\lambda_5x+\lambda_6y$ \\ $~\,\Phi=\lambda_1x+\lambda_2y+\lambda_3z$}  \\
\hline
\caption{Vector and scalar potentials of the electromagnetic field for four-parameter Noether symmetry groups generated by $\textbf{v}_0$, $\textbf{X}_1$, $\textbf{X}_2$ and $\textbf{X}_3$.}  
\label{nsym4d}
\end{longtable}
\renewcommand\arraystretch{1.0}

Higher-dimensional classification results would retrieve subcases of Table \ref{nsym3d} except for case 2 of Table \ref{nsym4d}, which does not contain any three-dimensional subalgebra. In the latter case, (where the magnetic field represents the magnetic monopole, $\boldsymbol{B}=\lambda\boldsymbol{r}/r^3$) the three Noether symmetry generators $\textbf{X}_1$, $\textbf{X}_2$ and $\textbf{X}_3$ correspond to the first integrals $\bar{I}_1=x\dot{y}-y\dot{x}-\lambda z/r$, $\bar{I}_2=z\dot{x}-x\dot{z}-\lambda y/r$ and $\bar{I}_3=y\dot{z}-z\dot{y}-\lambda x/r$, respectively. Following the structure of the symmetry algebra, they are not in involution though, $\left\{\bar{I}_1,\bar{I}_2\right\}=-\bar{I}_3$, $\left\{\bar{I}_2,\bar{I}_3\right\}=-\bar{I}_1$ and $\left\{\bar{I}_1,\bar{I}_3\right\}=\bar{I}_2$. Nevertheless, similar to the classical problem of central-force motion, taking $I_1=\bar{I}_1^2+\bar{I}_2^2+\bar{I}_3^2$ and any $\bar{I}_i$ as $I_2$, we have $\left\{I_1,I_2\right\}=\left\{\bar{I}_j\bar{I}_j,\bar{I}_i\right\}=2\bar{I}_j\left\{\bar{I}_j,\bar{I}_i\right\}=2\epsilon_{ijk}\bar{I}_j\bar{I}_k=0$. Note that in this case $I_1$ is quadratic in the velocities and corresponds to a contact symmetry generated by $2\bar{I}_i\textbf{X}_i$. By construction though either $I_1$ as a function of the integrals $\bar{I}_i$ that correspond to $\textbf{X}_i$, or $2\bar{I}_i\textbf{X}_i$ as a linear combination of $\textbf{X}_i$ with coefficients the corresponding integrals is really coming from the rotations generated by $\textbf{X}_i$, i.e. point symmetries. Under these considerations, we reach the following conclusion.

\begin{corollary}\normalfont
For inhomogeneous and curved magnetic fields, the autonomous system (\ref{primpotsystemeq}) of charged particle motion is completely integrable via Noether point symmetries in three representative cases under the equivalence transformations (\ref{group})\,:
\begin{enumerate}
	\item case 2 of Table \ref{nsym3d} for $\lambda_i=0$, where $I_1=\dot{z}+F_3$ and $I_2=x\dot{y}-y\dot{x}+(x^2+y^2)F_2$,
	\item case 5 of Table \ref{nsym3d} for $\lambda_i=0$, where $I_1=\dot{y}+F_2$ and $I_2=\dot{z}+F_3$,
	\item case 2 of Table \ref{nsym4d}, where $I_1=(x\dot{y}-y\dot{x})^2+(z\dot{x}-x\dot{z})^2+(y\dot{z}-z\dot{y})^2$ and $I_2=\bar{I}_i$ for any $\bar{I}_1=x\dot{y}-y\dot{x}-\lambda z/r$, $\bar{I}_2=z\dot{x}-x\dot{z}-\lambda y/r$, $\bar{I}_3=y\dot{z}-z\dot{y}-\lambda x/r$.
\end{enumerate}
\end{corollary}

\section{Discussion}

We have found and classified one-, two- and three-parameter symmetry group extensions of time translations admitted by the three-dimensional autonomous non-relativistic charged particle motion. The classification was made under the action of the equivalence group, which also proved to preserve the homogeneous Maxwell's equations with no restrictions at all. Therefore each symmetry class is described in terms of the vector and scalar potentials of the electromagnetic field in a representative form as simple as possible under equivalence transformations. In other words, the members that belong to each case can be found by transforming the typical potentials presented in the previous tables using (\ref{group}). The corresponding Noether symmetry classification has also led to some first conclusions about the existence of first integrals of motion besides the well-known Hamiltonian. 


In general, this type of classification is considered preliminary in the sense that there could be more symmetries lying outside the equivalence group. However, from the inspection of the determining equations resulting from the symmetry condition in \cite{me}, we concluded that, when the system is nonlinear and particularly when the magnetic field is inhomogeneous and curved, the only symmetries admitted are the ones that belong to the equivalence group. Therefore, in this case Tables \ref{sym2d}-\ref{sym4d} can be considered as a full classification in terms of point symmetries and up to three-parameter symmetry groups. Of course, more symmetries can be expected when the potentials listed in the previous tables result in magnetic fields of constant direction or even linear equations of motion.

Based on the results of this work, particular cases of physical interest can be further investigated by using symmetry methods to reduce the order of the system. The group classification of the non-autonomous case is naturally another desirable development and possibly more feasible, having ruled out the wide class of time-independent fields treated here.

\subsection*{Acknowledgements}
The author is gratefully indebted to Dr Stelios Dimas for valuable suggestions, useful comments and many helpful discussions.


\end{document}